\documentclass[11pt,a4paper]{article}
\pdfoutput=1
\usepackage{jcappub}
\usepackage{amsmath}
\usepackage{amssymb}

\usepackage{graphicx, subfig}
\usepackage{natbib}
\usepackage{aastex_hack}
\usepackage{color}
\definecolor{AliceBlue}{rgb}{0.94,0.97,1.00}
\definecolor{AntiqueWhite1}{rgb}{1.00,0.94,0.86}
\definecolor{AntiqueWhite2}{rgb}{0.93,0.87,0.80}
\definecolor{AntiqueWhite3}{rgb}{0.80,0.75,0.69}
\definecolor{AntiqueWhite4}{rgb}{0.55,0.51,0.47}
\definecolor{AntiqueWhite}{rgb}{0.98,0.92,0.84}
\definecolor{BlanchedAlmond}{rgb}{1.00,0.92,0.80}
\definecolor{BlueViolet}{rgb}{0.54,0.17,0.89}
\definecolor{CadetBlue1}{rgb}{0.60,0.96,1.00}
\definecolor{CadetBlue2}{rgb}{0.56,0.90,0.93}
\definecolor{CadetBlue3}{rgb}{0.48,0.77,0.80}
\definecolor{CadetBlue4}{rgb}{0.33,0.53,0.55}
\definecolor{CadetBlue}{rgb}{0.37,0.62,0.63}
\definecolor{CornflowerBlue}{rgb}{0.39,0.58,0.93}
\definecolor{DarkBlue}{rgb}{0.00,0.00,0.55}
\definecolor{DarkCyan}{rgb}{0.00,0.55,0.55}
\definecolor{DarkGoldenrod1}{rgb}{1.00,0.73,0.06}
\definecolor{DarkGoldenrod2}{rgb}{0.93,0.68,0.05}
\definecolor{DarkGoldenrod3}{rgb}{0.80,0.58,0.05}
\definecolor{DarkGoldenrod4}{rgb}{0.55,0.40,0.03}
\definecolor{DarkGoldenrod}{rgb}{0.72,0.53,0.04}
\definecolor{DarkGray}{rgb}{0.66,0.66,0.66}
\definecolor{DarkGreen}{rgb}{0.00,0.39,0.00}
\definecolor{DarkGrey}{rgb}{0.66,0.66,0.66}
\definecolor{DarkKhaki}{rgb}{0.74,0.72,0.42}
\definecolor{DarkMagenta}{rgb}{0.55,0.00,0.55}
\definecolor{DarkOliveGreen1}{rgb}{0.79,1.00,0.44}
\definecolor{DarkOliveGreen2}{rgb}{0.74,0.93,0.41}
\definecolor{DarkOliveGreen3}{rgb}{0.64,0.80,0.35}
\definecolor{DarkOliveGreen4}{rgb}{0.43,0.55,0.24}
\definecolor{DarkOliveGreen}{rgb}{0.33,0.42,0.18}
\definecolor{DarkOrange1}{rgb}{1.00,0.50,0.00}
\definecolor{DarkOrange2}{rgb}{0.93,0.46,0.00}
\definecolor{DarkOrange3}{rgb}{0.80,0.40,0.00}
\definecolor{DarkOrange4}{rgb}{0.55,0.27,0.00}
\definecolor{DarkOrange}{rgb}{1.00,0.55,0.00}
\definecolor{DarkOrchid1}{rgb}{0.75,0.24,1.00}
\definecolor{DarkOrchid2}{rgb}{0.70,0.23,0.93}
\definecolor{DarkOrchid3}{rgb}{0.60,0.20,0.80}
\definecolor{DarkOrchid4}{rgb}{0.41,0.13,0.55}
\definecolor{DarkOrchid}{rgb}{0.60,0.20,0.80}
\definecolor{DarkRed}{rgb}{0.55,0.00,0.00}
\definecolor{DarkSalmon}{rgb}{0.91,0.59,0.48}
\definecolor{DarkSeaGreen1}{rgb}{0.76,1.00,0.76}
\definecolor{DarkSeaGreen2}{rgb}{0.71,0.93,0.71}
\definecolor{DarkSeaGreen3}{rgb}{0.61,0.80,0.61}
\definecolor{DarkSeaGreen4}{rgb}{0.41,0.55,0.41}
\definecolor{DarkSeaGreen}{rgb}{0.56,0.74,0.56}
\definecolor{DarkSlateBlue}{rgb}{0.28,0.24,0.55}
\definecolor{DarkSlateGray1}{rgb}{0.59,1.00,1.00}
\definecolor{DarkSlateGray2}{rgb}{0.55,0.93,0.93}
\definecolor{DarkSlateGray3}{rgb}{0.47,0.80,0.80}
\definecolor{DarkSlateGray4}{rgb}{0.32,0.55,0.55}
\definecolor{DarkSlateGray}{rgb}{0.18,0.31,0.31}
\definecolor{DarkSlateGrey}{rgb}{0.18,0.31,0.31}
\definecolor{DarkTurquoise}{rgb}{0.00,0.81,0.82}
\definecolor{DarkViolet}{rgb}{0.58,0.00,0.83}
\definecolor{DeepPink1}{rgb}{1.00,0.08,0.58}
\definecolor{DeepPink2}{rgb}{0.93,0.07,0.54}
\definecolor{DeepPink3}{rgb}{0.80,0.06,0.46}
\definecolor{DeepPink4}{rgb}{0.55,0.04,0.31}
\definecolor{DeepPink}{rgb}{1.00,0.08,0.58}
\definecolor{DeepSkyBlue1}{rgb}{0.00,0.75,1.00}
\definecolor{DeepSkyBlue2}{rgb}{0.00,0.70,0.93}
\definecolor{DeepSkyBlue3}{rgb}{0.00,0.60,0.80}
\definecolor{DeepSkyBlue4}{rgb}{0.00,0.41,0.55}
\definecolor{DeepSkyBlue}{rgb}{0.00,0.75,1.00}
\definecolor{DimGray}{rgb}{0.41,0.41,0.41}
\definecolor{DimGrey}{rgb}{0.41,0.41,0.41}
\definecolor{DodgerBlue1}{rgb}{0.12,0.56,1.00}
\definecolor{DodgerBlue2}{rgb}{0.11,0.53,0.93}
\definecolor{DodgerBlue3}{rgb}{0.09,0.45,0.80}
\definecolor{DodgerBlue4}{rgb}{0.06,0.31,0.55}
\definecolor{DodgerBlue}{rgb}{0.12,0.56,1.00}
\definecolor{FloralWhite}{rgb}{1.00,0.98,0.94}
\definecolor{ForestGreen}{rgb}{0.13,0.55,0.13}
\definecolor{GhostWhite}{rgb}{0.97,0.97,1.00}
\definecolor{GreenYellow}{rgb}{0.68,1.00,0.18}
\definecolor{HotPink1}{rgb}{1.00,0.43,0.71}
\definecolor{HotPink2}{rgb}{0.93,0.42,0.65}
\definecolor{HotPink3}{rgb}{0.80,0.38,0.56}
\definecolor{HotPink4}{rgb}{0.55,0.23,0.38}
\definecolor{HotPink}{rgb}{1.00,0.41,0.71}
\definecolor{IndianRed1}{rgb}{1.00,0.42,0.42}
\definecolor{IndianRed2}{rgb}{0.93,0.39,0.39}
\definecolor{IndianRed3}{rgb}{0.80,0.33,0.33}
\definecolor{IndianRed4}{rgb}{0.55,0.23,0.23}
\definecolor{IndianRed}{rgb}{0.80,0.36,0.36}
\definecolor{LavenderBlush1}{rgb}{1.00,0.94,0.96}
\definecolor{LavenderBlush2}{rgb}{0.93,0.88,0.90}
\definecolor{LavenderBlush3}{rgb}{0.80,0.76,0.77}
\definecolor{LavenderBlush4}{rgb}{0.55,0.51,0.53}
\definecolor{LavenderBlush}{rgb}{1.00,0.94,0.96}
\definecolor{LawnGreen}{rgb}{0.49,0.99,0.00}
\definecolor{LemonChiffon1}{rgb}{1.00,0.98,0.80}
\definecolor{LemonChiffon2}{rgb}{0.93,0.91,0.75}
\definecolor{LemonChiffon3}{rgb}{0.80,0.79,0.65}
\definecolor{LemonChiffon4}{rgb}{0.55,0.54,0.44}
\definecolor{LemonChiffon}{rgb}{1.00,0.98,0.80}
\definecolor{LightBlue1}{rgb}{0.75,0.94,1.00}
\definecolor{LightBlue2}{rgb}{0.70,0.87,0.93}
\definecolor{LightBlue3}{rgb}{0.60,0.75,0.80}
\definecolor{LightBlue4}{rgb}{0.41,0.51,0.55}
\definecolor{LightBlue}{rgb}{0.68,0.85,0.90}
\definecolor{LightCoral}{rgb}{0.94,0.50,0.50}
\definecolor{LightCyan1}{rgb}{0.88,1.00,1.00}
\definecolor{LightCyan2}{rgb}{0.82,0.93,0.93}
\definecolor{LightCyan3}{rgb}{0.71,0.80,0.80}
\definecolor{LightCyan4}{rgb}{0.48,0.55,0.55}
\definecolor{LightCyan}{rgb}{0.88,1.00,1.00}
\definecolor{LightGoldenrod1}{rgb}{1.00,0.93,0.55}
\definecolor{LightGoldenrod2}{rgb}{0.93,0.86,0.51}
\definecolor{LightGoldenrod3}{rgb}{0.80,0.75,0.44}
\definecolor{LightGoldenrod4}{rgb}{0.55,0.51,0.30}
\definecolor{LightGoldenrodYellow}{rgb}{0.98,0.98,0.82}
\definecolor{LightGoldenrod}{rgb}{0.93,0.87,0.51}
\definecolor{LightGray}{rgb}{0.83,0.83,0.83}
\definecolor{LightGreen}{rgb}{0.56,0.93,0.56}
\definecolor{LightGrey}{rgb}{0.83,0.83,0.83}
\definecolor{LightPink1}{rgb}{1.00,0.68,0.73}
\definecolor{LightPink2}{rgb}{0.93,0.64,0.68}
\definecolor{LightPink3}{rgb}{0.80,0.55,0.58}
\definecolor{LightPink4}{rgb}{0.55,0.37,0.40}
\definecolor{LightPink}{rgb}{1.00,0.71,0.76}
\definecolor{LightSalmon1}{rgb}{1.00,0.63,0.48}
\definecolor{LightSalmon2}{rgb}{0.93,0.58,0.45}
\definecolor{LightSalmon3}{rgb}{0.80,0.51,0.38}
\definecolor{LightSalmon4}{rgb}{0.55,0.34,0.26}
\definecolor{LightSalmon}{rgb}{1.00,0.63,0.48}
\definecolor{LightSeaGreen}{rgb}{0.13,0.70,0.67}
\definecolor{LightSkyBlue1}{rgb}{0.69,0.89,1.00}
\definecolor{LightSkyBlue2}{rgb}{0.64,0.83,0.93}
\definecolor{LightSkyBlue3}{rgb}{0.55,0.71,0.80}
\definecolor{LightSkyBlue4}{rgb}{0.38,0.48,0.55}
\definecolor{LightSkyBlue}{rgb}{0.53,0.81,0.98}
\definecolor{LightSlateBlue}{rgb}{0.52,0.44,1.00}
\definecolor{LightSlateGray}{rgb}{0.47,0.53,0.60}
\definecolor{LightSlateGrey}{rgb}{0.47,0.53,0.60}
\definecolor{LightSteelBlue1}{rgb}{0.79,0.88,1.00}
\definecolor{LightSteelBlue2}{rgb}{0.74,0.82,0.93}
\definecolor{LightSteelBlue3}{rgb}{0.64,0.71,0.80}
\definecolor{LightSteelBlue4}{rgb}{0.43,0.48,0.55}
\definecolor{LightSteelBlue}{rgb}{0.69,0.77,0.87}
\definecolor{LightYellow1}{rgb}{1.00,1.00,0.88}
\definecolor{LightYellow2}{rgb}{0.93,0.93,0.82}
\definecolor{LightYellow3}{rgb}{0.80,0.80,0.71}
\definecolor{LightYellow4}{rgb}{0.55,0.55,0.48}
\definecolor{LightYellow}{rgb}{1.00,1.00,0.88}
\definecolor{LimeGreen}{rgb}{0.20,0.80,0.20}
\definecolor{MediumAquamarine}{rgb}{0.40,0.80,0.67}
\definecolor{MediumBlue}{rgb}{0.00,0.00,0.80}
\definecolor{MediumOrchid1}{rgb}{0.88,0.40,1.00}
\definecolor{MediumOrchid2}{rgb}{0.82,0.37,0.93}
\definecolor{MediumOrchid3}{rgb}{0.71,0.32,0.80}
\definecolor{MediumOrchid4}{rgb}{0.48,0.22,0.55}
\definecolor{MediumOrchid}{rgb}{0.73,0.33,0.83}
\definecolor{MediumPurple1}{rgb}{0.67,0.51,1.00}
\definecolor{MediumPurple2}{rgb}{0.62,0.47,0.93}
\definecolor{MediumPurple3}{rgb}{0.54,0.41,0.80}
\definecolor{MediumPurple4}{rgb}{0.36,0.28,0.55}
\definecolor{MediumPurple}{rgb}{0.58,0.44,0.86}
\definecolor{MediumSeaGreen}{rgb}{0.24,0.70,0.44}
\definecolor{MediumSlateBlue}{rgb}{0.48,0.41,0.93}
\definecolor{MediumSpringGreen}{rgb}{0.00,0.98,0.60}
\definecolor{MediumTurquoise}{rgb}{0.28,0.82,0.80}
\definecolor{MediumVioletRed}{rgb}{0.78,0.08,0.52}
\definecolor{MidnightBlue}{rgb}{0.10,0.10,0.44}
\definecolor{MintCream}{rgb}{0.96,1.00,0.98}
\definecolor{MistyRose1}{rgb}{1.00,0.89,0.88}
\definecolor{MistyRose2}{rgb}{0.93,0.84,0.82}
\definecolor{MistyRose3}{rgb}{0.80,0.72,0.71}
\definecolor{MistyRose4}{rgb}{0.55,0.49,0.48}
\definecolor{MistyRose}{rgb}{1.00,0.89,0.88}
\definecolor{NavajoWhite1}{rgb}{1.00,0.87,0.68}
\definecolor{NavajoWhite2}{rgb}{0.93,0.81,0.63}
\definecolor{NavajoWhite3}{rgb}{0.80,0.70,0.55}
\definecolor{NavajoWhite4}{rgb}{0.55,0.47,0.37}
\definecolor{NavajoWhite}{rgb}{1.00,0.87,0.68}
\definecolor{NavyBlue}{rgb}{0.00,0.00,0.50}
\definecolor{OldLace}{rgb}{0.99,0.96,0.90}
\definecolor{OliveDrab1}{rgb}{0.75,1.00,0.24}
\definecolor{OliveDrab2}{rgb}{0.70,0.93,0.23}
\definecolor{OliveDrab3}{rgb}{0.60,0.80,0.20}
\definecolor{OliveDrab4}{rgb}{0.41,0.55,0.13}
\definecolor{OliveDrab}{rgb}{0.42,0.56,0.14}
\definecolor{OrangeRed1}{rgb}{1.00,0.27,0.00}
\definecolor{OrangeRed2}{rgb}{0.93,0.25,0.00}
\definecolor{OrangeRed3}{rgb}{0.80,0.22,0.00}
\definecolor{OrangeRed4}{rgb}{0.55,0.15,0.00}
\definecolor{OrangeRed}{rgb}{1.00,0.27,0.00}
\definecolor{PaleGoldenrod}{rgb}{0.93,0.91,0.67}
\definecolor{PaleGreen1}{rgb}{0.60,1.00,0.60}
\definecolor{PaleGreen2}{rgb}{0.56,0.93,0.56}
\definecolor{PaleGreen3}{rgb}{0.49,0.80,0.49}
\definecolor{PaleGreen4}{rgb}{0.33,0.55,0.33}
\definecolor{PaleGreen}{rgb}{0.60,0.98,0.60}
\definecolor{PaleTurquoise1}{rgb}{0.73,1.00,1.00}
\definecolor{PaleTurquoise2}{rgb}{0.68,0.93,0.93}
\definecolor{PaleTurquoise3}{rgb}{0.59,0.80,0.80}
\definecolor{PaleTurquoise4}{rgb}{0.40,0.55,0.55}
\definecolor{PaleTurquoise}{rgb}{0.69,0.93,0.93}
\definecolor{PaleVioletRed1}{rgb}{1.00,0.51,0.67}
\definecolor{PaleVioletRed2}{rgb}{0.93,0.47,0.62}
\definecolor{PaleVioletRed3}{rgb}{0.80,0.41,0.54}
\definecolor{PaleVioletRed4}{rgb}{0.55,0.28,0.36}
\definecolor{PaleVioletRed}{rgb}{0.86,0.44,0.58}
\definecolor{PapayaWhip}{rgb}{1.00,0.94,0.84}
\definecolor{PeachPuff1}{rgb}{1.00,0.85,0.73}
\definecolor{PeachPuff2}{rgb}{0.93,0.80,0.68}
\definecolor{PeachPuff3}{rgb}{0.80,0.69,0.58}
\definecolor{PeachPuff4}{rgb}{0.55,0.47,0.40}
\definecolor{PeachPuff}{rgb}{1.00,0.85,0.73}
\definecolor{PowderBlue}{rgb}{0.69,0.88,0.90}
\definecolor{RosyBrown1}{rgb}{1.00,0.76,0.76}
\definecolor{RosyBrown2}{rgb}{0.93,0.71,0.71}
\definecolor{RosyBrown3}{rgb}{0.80,0.61,0.61}
\definecolor{RosyBrown4}{rgb}{0.55,0.41,0.41}
\definecolor{RosyBrown}{rgb}{0.74,0.56,0.56}
\definecolor{RoyalBlue1}{rgb}{0.28,0.46,1.00}
\definecolor{RoyalBlue2}{rgb}{0.26,0.43,0.93}
\definecolor{RoyalBlue3}{rgb}{0.23,0.37,0.80}
\definecolor{RoyalBlue4}{rgb}{0.15,0.25,0.55}
\definecolor{RoyalBlue}{rgb}{0.25,0.41,0.88}
\definecolor{SaddleBrown}{rgb}{0.55,0.27,0.07}
\definecolor{SandyBrown}{rgb}{0.96,0.64,0.38}
\definecolor{SeaGreen1}{rgb}{0.33,1.00,0.62}
\definecolor{SeaGreen2}{rgb}{0.31,0.93,0.58}
\definecolor{SeaGreen3}{rgb}{0.26,0.80,0.50}
\definecolor{SeaGreen4}{rgb}{0.18,0.55,0.34}
\definecolor{SeaGreen}{rgb}{0.18,0.55,0.34}
\definecolor{SkyBlue1}{rgb}{0.53,0.81,1.00}
\definecolor{SkyBlue2}{rgb}{0.49,0.75,0.93}
\definecolor{SkyBlue3}{rgb}{0.42,0.65,0.80}
\definecolor{SkyBlue4}{rgb}{0.29,0.44,0.55}
\definecolor{SkyBlue}{rgb}{0.53,0.81,0.92}
\definecolor{SlateBlue1}{rgb}{0.51,0.44,1.00}
\definecolor{SlateBlue2}{rgb}{0.48,0.40,0.93}
\definecolor{SlateBlue3}{rgb}{0.41,0.35,0.80}
\definecolor{SlateBlue4}{rgb}{0.28,0.24,0.55}
\definecolor{SlateBlue}{rgb}{0.42,0.35,0.80}
\definecolor{SlateGray1}{rgb}{0.78,0.89,1.00}
\definecolor{SlateGray2}{rgb}{0.73,0.83,0.93}
\definecolor{SlateGray3}{rgb}{0.62,0.71,0.80}
\definecolor{SlateGray4}{rgb}{0.42,0.48,0.55}
\definecolor{SlateGray}{rgb}{0.44,0.50,0.56}
\definecolor{SlateGrey}{rgb}{0.44,0.50,0.56}
\definecolor{SpringGreen1}{rgb}{0.00,1.00,0.50}
\definecolor{SpringGreen2}{rgb}{0.00,0.93,0.46}
\definecolor{SpringGreen3}{rgb}{0.00,0.80,0.40}
\definecolor{SpringGreen4}{rgb}{0.00,0.55,0.27}
\definecolor{SpringGreen}{rgb}{0.00,1.00,0.50}
\definecolor{SteelBlue1}{rgb}{0.39,0.72,1.00}
\definecolor{SteelBlue2}{rgb}{0.36,0.67,0.93}
\definecolor{SteelBlue3}{rgb}{0.31,0.58,0.80}
\definecolor{SteelBlue4}{rgb}{0.21,0.39,0.55}
\definecolor{SteelBlue}{rgb}{0.27,0.51,0.71}
\definecolor{VioletRed1}{rgb}{1.00,0.24,0.59}
\definecolor{VioletRed2}{rgb}{0.93,0.23,0.55}
\definecolor{VioletRed3}{rgb}{0.80,0.20,0.47}
\definecolor{VioletRed4}{rgb}{0.55,0.13,0.32}
\definecolor{VioletRed}{rgb}{0.82,0.13,0.56}
\definecolor{WhiteSmoke}{rgb}{0.96,0.96,0.96}
\definecolor{YellowGreen}{rgb}{0.60,0.80,0.20}
\definecolor{aliceblue}{rgb}{0.94,0.97,1.00}
\definecolor{antiquewhite}{rgb}{0.98,0.92,0.84}
\definecolor{aquamarine1}{rgb}{0.50,1.00,0.83}
\definecolor{aquamarine2}{rgb}{0.46,0.93,0.78}
\definecolor{aquamarine3}{rgb}{0.40,0.80,0.67}
\definecolor{aquamarine4}{rgb}{0.27,0.55,0.45}
\definecolor{aquamarine}{rgb}{0.50,1.00,0.83}
\definecolor{azure1}{rgb}{0.94,1.00,1.00}
\definecolor{azure2}{rgb}{0.88,0.93,0.93}
\definecolor{azure3}{rgb}{0.76,0.80,0.80}
\definecolor{azure4}{rgb}{0.51,0.55,0.55}
\definecolor{azure}{rgb}{0.94,1.00,1.00}
\definecolor{beige}{rgb}{0.96,0.96,0.86}
\definecolor{bisque1}{rgb}{1.00,0.89,0.77}
\definecolor{bisque2}{rgb}{0.93,0.84,0.72}
\definecolor{bisque3}{rgb}{0.80,0.72,0.62}
\definecolor{bisque4}{rgb}{0.55,0.49,0.42}
\definecolor{bisque}{rgb}{1.00,0.89,0.77}
\definecolor{black}{rgb}{0.00,0.00,0.00}
\definecolor{blanchedalmond}{rgb}{1.00,0.92,0.80}
\definecolor{blue1}{rgb}{0.00,0.00,1.00}
\definecolor{blue2}{rgb}{0.00,0.00,0.93}
\definecolor{blue3}{rgb}{0.00,0.00,0.80}
\definecolor{blue4}{rgb}{0.00,0.00,0.55}
\definecolor{blueviolet}{rgb}{0.54,0.17,0.89}
\definecolor{blue}{rgb}{0.00,0.00,1.00}
\definecolor{brown1}{rgb}{1.00,0.25,0.25}
\definecolor{brown2}{rgb}{0.93,0.23,0.23}
\definecolor{brown3}{rgb}{0.80,0.20,0.20}
\definecolor{brown4}{rgb}{0.55,0.14,0.14}
\definecolor{brown}{rgb}{0.65,0.16,0.16}
\definecolor{burlywood1}{rgb}{1.00,0.83,0.61}
\definecolor{burlywood2}{rgb}{0.93,0.77,0.57}
\definecolor{burlywood3}{rgb}{0.80,0.67,0.49}
\definecolor{burlywood4}{rgb}{0.55,0.45,0.33}
\definecolor{burlywood}{rgb}{0.87,0.72,0.53}
\definecolor{cadetblue}{rgb}{0.37,0.62,0.63}
\definecolor{chartreuse1}{rgb}{0.50,1.00,0.00}
\definecolor{chartreuse2}{rgb}{0.46,0.93,0.00}
\definecolor{chartreuse3}{rgb}{0.40,0.80,0.00}
\definecolor{chartreuse4}{rgb}{0.27,0.55,0.00}
\definecolor{chartreuse}{rgb}{0.50,1.00,0.00}
\definecolor{chocolate1}{rgb}{1.00,0.50,0.14}
\definecolor{chocolate2}{rgb}{0.93,0.46,0.13}
\definecolor{chocolate3}{rgb}{0.80,0.40,0.11}
\definecolor{chocolate4}{rgb}{0.55,0.27,0.07}
\definecolor{chocolate}{rgb}{0.82,0.41,0.12}
\definecolor{coral1}{rgb}{1.00,0.45,0.34}
\definecolor{coral2}{rgb}{0.93,0.42,0.31}
\definecolor{coral3}{rgb}{0.80,0.36,0.27}
\definecolor{coral4}{rgb}{0.55,0.24,0.18}
\definecolor{coral}{rgb}{1.00,0.50,0.31}
\definecolor{cornflowerblue}{rgb}{0.39,0.58,0.93}
\definecolor{cornsilk1}{rgb}{1.00,0.97,0.86}
\definecolor{cornsilk2}{rgb}{0.93,0.91,0.80}
\definecolor{cornsilk3}{rgb}{0.80,0.78,0.69}
\definecolor{cornsilk4}{rgb}{0.55,0.53,0.47}
\definecolor{cornsilk}{rgb}{1.00,0.97,0.86}
\definecolor{cyan1}{rgb}{0.00,1.00,1.00}
\definecolor{cyan2}{rgb}{0.00,0.93,0.93}
\definecolor{cyan3}{rgb}{0.00,0.80,0.80}
\definecolor{cyan4}{rgb}{0.00,0.55,0.55}
\definecolor{cyan}{rgb}{0.00,1.00,1.00}
\definecolor{darkblue}{rgb}{0.00,0.00,0.55}
\definecolor{darkcyan}{rgb}{0.00,0.55,0.55}
\definecolor{darkgoldenrod}{rgb}{0.72,0.53,0.04}
\definecolor{darkgray}{rgb}{0.66,0.66,0.66}
\definecolor{darkgreen}{rgb}{0.00,0.39,0.00}
\definecolor{darkgrey}{rgb}{0.66,0.66,0.66}
\definecolor{darkkhaki}{rgb}{0.74,0.72,0.42}
\definecolor{darkmagenta}{rgb}{0.55,0.00,0.55}
\definecolor{darkolive}{rgb}{0.33,0.42,0.18}
\definecolor{darkorange}{rgb}{1.00,0.55,0.00}
\definecolor{darkorchid}{rgb}{0.60,0.20,0.80}
\definecolor{darkred}{rgb}{0.55,0.00,0.00}
\definecolor{darksalmon}{rgb}{0.91,0.59,0.48}
\definecolor{darksea}{rgb}{0.56,0.74,0.56}
\definecolor{darkslate}{rgb}{0.18,0.31,0.31}
\definecolor{darkslate}{rgb}{0.18,0.31,0.31}
\definecolor{darkslate}{rgb}{0.28,0.24,0.55}
\definecolor{darkturquoise}{rgb}{0.00,0.81,0.82}
\definecolor{darkviolet}{rgb}{0.58,0.00,0.83}
\definecolor{deeppink}{rgb}{1.00,0.08,0.58}
\definecolor{deepsky}{rgb}{0.00,0.75,1.00}
\definecolor{dimgray}{rgb}{0.41,0.41,0.41}
\definecolor{dimgrey}{rgb}{0.41,0.41,0.41}
\definecolor{dodgerblue}{rgb}{0.12,0.56,1.00}
\definecolor{firebrick1}{rgb}{1.00,0.19,0.19}
\definecolor{firebrick2}{rgb}{0.93,0.17,0.17}
\definecolor{firebrick3}{rgb}{0.80,0.15,0.15}
\definecolor{firebrick4}{rgb}{0.55,0.10,0.10}
\definecolor{firebrick}{rgb}{0.70,0.13,0.13}
\definecolor{floralwhite}{rgb}{1.00,0.98,0.94}
\definecolor{forestgreen}{rgb}{0.13,0.55,0.13}
\definecolor{gainsboro}{rgb}{0.86,0.86,0.86}
\definecolor{ghostwhite}{rgb}{0.97,0.97,1.00}
\definecolor{gold1}{rgb}{1.00,0.84,0.00}
\definecolor{gold2}{rgb}{0.93,0.79,0.00}
\definecolor{gold3}{rgb}{0.80,0.68,0.00}
\definecolor{gold4}{rgb}{0.55,0.46,0.00}
\definecolor{goldenrod1}{rgb}{1.00,0.76,0.15}
\definecolor{goldenrod2}{rgb}{0.93,0.71,0.13}
\definecolor{goldenrod3}{rgb}{0.80,0.61,0.11}
\definecolor{goldenrod4}{rgb}{0.55,0.41,0.08}
\definecolor{goldenrod}{rgb}{0.85,0.65,0.13}
\definecolor{gold}{rgb}{1.00,0.84,0.00}
\definecolor{gray0}{rgb}{0.00,0.00,0.00}
\definecolor{gray100}{rgb}{1.00,1.00,1.00}
\definecolor{gray10}{rgb}{0.10,0.10,0.10}
\definecolor{gray11}{rgb}{0.11,0.11,0.11}
\definecolor{gray12}{rgb}{0.12,0.12,0.12}
\definecolor{gray13}{rgb}{0.13,0.13,0.13}
\definecolor{gray14}{rgb}{0.14,0.14,0.14}
\definecolor{gray15}{rgb}{0.15,0.15,0.15}
\definecolor{gray16}{rgb}{0.16,0.16,0.16}
\definecolor{gray17}{rgb}{0.17,0.17,0.17}
\definecolor{gray18}{rgb}{0.18,0.18,0.18}
\definecolor{gray19}{rgb}{0.19,0.19,0.19}
\definecolor{gray1}{rgb}{0.01,0.01,0.01}
\definecolor{gray20}{rgb}{0.20,0.20,0.20}
\definecolor{gray21}{rgb}{0.21,0.21,0.21}
\definecolor{gray22}{rgb}{0.22,0.22,0.22}
\definecolor{gray23}{rgb}{0.23,0.23,0.23}
\definecolor{gray24}{rgb}{0.24,0.24,0.24}
\definecolor{gray25}{rgb}{0.25,0.25,0.25}
\definecolor{gray26}{rgb}{0.26,0.26,0.26}
\definecolor{gray27}{rgb}{0.27,0.27,0.27}
\definecolor{gray28}{rgb}{0.28,0.28,0.28}
\definecolor{gray29}{rgb}{0.29,0.29,0.29}
\definecolor{gray2}{rgb}{0.02,0.02,0.02}
\definecolor{gray30}{rgb}{0.30,0.30,0.30}
\definecolor{gray31}{rgb}{0.31,0.31,0.31}
\definecolor{gray32}{rgb}{0.32,0.32,0.32}
\definecolor{gray33}{rgb}{0.33,0.33,0.33}
\definecolor{gray34}{rgb}{0.34,0.34,0.34}
\definecolor{gray35}{rgb}{0.35,0.35,0.35}
\definecolor{gray36}{rgb}{0.36,0.36,0.36}
\definecolor{gray37}{rgb}{0.37,0.37,0.37}
\definecolor{gray38}{rgb}{0.38,0.38,0.38}
\definecolor{gray39}{rgb}{0.39,0.39,0.39}
\definecolor{gray3}{rgb}{0.03,0.03,0.03}
\definecolor{gray40}{rgb}{0.40,0.40,0.40}
\definecolor{gray41}{rgb}{0.41,0.41,0.41}
\definecolor{gray42}{rgb}{0.42,0.42,0.42}
\definecolor{gray43}{rgb}{0.43,0.43,0.43}
\definecolor{gray44}{rgb}{0.44,0.44,0.44}
\definecolor{gray45}{rgb}{0.45,0.45,0.45}
\definecolor{gray46}{rgb}{0.46,0.46,0.46}
\definecolor{gray47}{rgb}{0.47,0.47,0.47}
\definecolor{gray48}{rgb}{0.48,0.48,0.48}
\definecolor{gray49}{rgb}{0.49,0.49,0.49}
\definecolor{gray4}{rgb}{0.04,0.04,0.04}
\definecolor{gray50}{rgb}{0.50,0.50,0.50}
\definecolor{gray51}{rgb}{0.51,0.51,0.51}
\definecolor{gray52}{rgb}{0.52,0.52,0.52}
\definecolor{gray53}{rgb}{0.53,0.53,0.53}
\definecolor{gray54}{rgb}{0.54,0.54,0.54}
\definecolor{gray55}{rgb}{0.55,0.55,0.55}
\definecolor{gray56}{rgb}{0.56,0.56,0.56}
\definecolor{gray57}{rgb}{0.57,0.57,0.57}
\definecolor{gray58}{rgb}{0.58,0.58,0.58}
\definecolor{gray59}{rgb}{0.59,0.59,0.59}
\definecolor{gray5}{rgb}{0.05,0.05,0.05}
\definecolor{gray60}{rgb}{0.60,0.60,0.60}
\definecolor{gray61}{rgb}{0.61,0.61,0.61}
\definecolor{gray62}{rgb}{0.62,0.62,0.62}
\definecolor{gray63}{rgb}{0.63,0.63,0.63}
\definecolor{gray64}{rgb}{0.64,0.64,0.64}
\definecolor{gray65}{rgb}{0.65,0.65,0.65}
\definecolor{gray66}{rgb}{0.66,0.66,0.66}
\definecolor{gray67}{rgb}{0.67,0.67,0.67}
\definecolor{gray68}{rgb}{0.68,0.68,0.68}
\definecolor{gray69}{rgb}{0.69,0.69,0.69}
\definecolor{gray6}{rgb}{0.06,0.06,0.06}
\definecolor{gray70}{rgb}{0.70,0.70,0.70}
\definecolor{gray71}{rgb}{0.71,0.71,0.71}
\definecolor{gray72}{rgb}{0.72,0.72,0.72}
\definecolor{gray73}{rgb}{0.73,0.73,0.73}
\definecolor{gray74}{rgb}{0.74,0.74,0.74}
\definecolor{gray75}{rgb}{0.75,0.75,0.75}
\definecolor{gray76}{rgb}{0.76,0.76,0.76}
\definecolor{gray77}{rgb}{0.77,0.77,0.77}
\definecolor{gray78}{rgb}{0.78,0.78,0.78}
\definecolor{gray79}{rgb}{0.79,0.79,0.79}
\definecolor{gray7}{rgb}{0.07,0.07,0.07}
\definecolor{gray80}{rgb}{0.80,0.80,0.80}
\definecolor{gray81}{rgb}{0.81,0.81,0.81}
\definecolor{gray82}{rgb}{0.82,0.82,0.82}
\definecolor{gray83}{rgb}{0.83,0.83,0.83}
\definecolor{gray84}{rgb}{0.84,0.84,0.84}
\definecolor{gray85}{rgb}{0.85,0.85,0.85}
\definecolor{gray86}{rgb}{0.86,0.86,0.86}
\definecolor{gray87}{rgb}{0.87,0.87,0.87}
\definecolor{gray88}{rgb}{0.88,0.88,0.88}
\definecolor{gray89}{rgb}{0.89,0.89,0.89}
\definecolor{gray8}{rgb}{0.08,0.08,0.08}
\definecolor{gray90}{rgb}{0.90,0.90,0.90}
\definecolor{gray91}{rgb}{0.91,0.91,0.91}
\definecolor{gray92}{rgb}{0.92,0.92,0.92}
\definecolor{gray93}{rgb}{0.93,0.93,0.93}
\definecolor{gray94}{rgb}{0.94,0.94,0.94}
\definecolor{gray95}{rgb}{0.95,0.95,0.95}
\definecolor{gray96}{rgb}{0.96,0.96,0.96}
\definecolor{gray97}{rgb}{0.97,0.97,0.97}
\definecolor{gray98}{rgb}{0.98,0.98,0.98}
\definecolor{gray99}{rgb}{0.99,0.99,0.99}
\definecolor{gray9}{rgb}{0.09,0.09,0.09}
\definecolor{gray}{rgb}{0.75,0.75,0.75}
\definecolor{green1}{rgb}{0.00,1.00,0.00}
\definecolor{green2}{rgb}{0.00,0.93,0.00}
\definecolor{green3}{rgb}{0.00,0.80,0.00}
\definecolor{green4}{rgb}{0.00,0.55,0.00}
\definecolor{greenyellow}{rgb}{0.68,1.00,0.18}
\definecolor{green}{rgb}{0.00,1.00,0.00}
\definecolor{grey0}{rgb}{0.00,0.00,0.00}
\definecolor{grey100}{rgb}{1.00,1.00,1.00}
\definecolor{grey10}{rgb}{0.10,0.10,0.10}
\definecolor{grey11}{rgb}{0.11,0.11,0.11}
\definecolor{grey12}{rgb}{0.12,0.12,0.12}
\definecolor{grey13}{rgb}{0.13,0.13,0.13}
\definecolor{grey14}{rgb}{0.14,0.14,0.14}
\definecolor{grey15}{rgb}{0.15,0.15,0.15}
\definecolor{grey16}{rgb}{0.16,0.16,0.16}
\definecolor{grey17}{rgb}{0.17,0.17,0.17}
\definecolor{grey18}{rgb}{0.18,0.18,0.18}
\definecolor{grey19}{rgb}{0.19,0.19,0.19}
\definecolor{grey1}{rgb}{0.01,0.01,0.01}
\definecolor{grey20}{rgb}{0.20,0.20,0.20}
\definecolor{grey21}{rgb}{0.21,0.21,0.21}
\definecolor{grey22}{rgb}{0.22,0.22,0.22}
\definecolor{grey23}{rgb}{0.23,0.23,0.23}
\definecolor{grey24}{rgb}{0.24,0.24,0.24}
\definecolor{grey25}{rgb}{0.25,0.25,0.25}
\definecolor{grey26}{rgb}{0.26,0.26,0.26}
\definecolor{grey27}{rgb}{0.27,0.27,0.27}
\definecolor{grey28}{rgb}{0.28,0.28,0.28}
\definecolor{grey29}{rgb}{0.29,0.29,0.29}
\definecolor{grey2}{rgb}{0.02,0.02,0.02}
\definecolor{grey30}{rgb}{0.30,0.30,0.30}
\definecolor{grey31}{rgb}{0.31,0.31,0.31}
\definecolor{grey32}{rgb}{0.32,0.32,0.32}
\definecolor{grey33}{rgb}{0.33,0.33,0.33}
\definecolor{grey34}{rgb}{0.34,0.34,0.34}
\definecolor{grey35}{rgb}{0.35,0.35,0.35}
\definecolor{grey36}{rgb}{0.36,0.36,0.36}
\definecolor{grey37}{rgb}{0.37,0.37,0.37}
\definecolor{grey38}{rgb}{0.38,0.38,0.38}
\definecolor{grey39}{rgb}{0.39,0.39,0.39}
\definecolor{grey3}{rgb}{0.03,0.03,0.03}
\definecolor{grey40}{rgb}{0.40,0.40,0.40}
\definecolor{grey41}{rgb}{0.41,0.41,0.41}
\definecolor{grey42}{rgb}{0.42,0.42,0.42}
\definecolor{grey43}{rgb}{0.43,0.43,0.43}
\definecolor{grey44}{rgb}{0.44,0.44,0.44}
\definecolor{grey45}{rgb}{0.45,0.45,0.45}
\definecolor{grey46}{rgb}{0.46,0.46,0.46}
\definecolor{grey47}{rgb}{0.47,0.47,0.47}
\definecolor{grey48}{rgb}{0.48,0.48,0.48}
\definecolor{grey49}{rgb}{0.49,0.49,0.49}
\definecolor{grey4}{rgb}{0.04,0.04,0.04}
\definecolor{grey50}{rgb}{0.50,0.50,0.50}
\definecolor{grey51}{rgb}{0.51,0.51,0.51}
\definecolor{grey52}{rgb}{0.52,0.52,0.52}
\definecolor{grey53}{rgb}{0.53,0.53,0.53}
\definecolor{grey54}{rgb}{0.54,0.54,0.54}
\definecolor{grey55}{rgb}{0.55,0.55,0.55}
\definecolor{grey56}{rgb}{0.56,0.56,0.56}
\definecolor{grey57}{rgb}{0.57,0.57,0.57}
\definecolor{grey58}{rgb}{0.58,0.58,0.58}
\definecolor{grey59}{rgb}{0.59,0.59,0.59}
\definecolor{grey5}{rgb}{0.05,0.05,0.05}
\definecolor{grey60}{rgb}{0.60,0.60,0.60}
\definecolor{grey61}{rgb}{0.61,0.61,0.61}
\definecolor{grey62}{rgb}{0.62,0.62,0.62}
\definecolor{grey63}{rgb}{0.63,0.63,0.63}
\definecolor{grey64}{rgb}{0.64,0.64,0.64}
\definecolor{grey65}{rgb}{0.65,0.65,0.65}
\definecolor{grey66}{rgb}{0.66,0.66,0.66}
\definecolor{grey67}{rgb}{0.67,0.67,0.67}
\definecolor{grey68}{rgb}{0.68,0.68,0.68}
\definecolor{grey69}{rgb}{0.69,0.69,0.69}
\definecolor{grey6}{rgb}{0.06,0.06,0.06}
\definecolor{grey70}{rgb}{0.70,0.70,0.70}
\definecolor{grey71}{rgb}{0.71,0.71,0.71}
\definecolor{grey72}{rgb}{0.72,0.72,0.72}
\definecolor{grey73}{rgb}{0.73,0.73,0.73}
\definecolor{grey74}{rgb}{0.74,0.74,0.74}
\definecolor{grey75}{rgb}{0.75,0.75,0.75}
\definecolor{grey76}{rgb}{0.76,0.76,0.76}
\definecolor{grey77}{rgb}{0.77,0.77,0.77}
\definecolor{grey78}{rgb}{0.78,0.78,0.78}
\definecolor{grey79}{rgb}{0.79,0.79,0.79}
\definecolor{grey7}{rgb}{0.07,0.07,0.07}
\definecolor{grey80}{rgb}{0.80,0.80,0.80}
\definecolor{grey81}{rgb}{0.81,0.81,0.81}
\definecolor{grey82}{rgb}{0.82,0.82,0.82}
\definecolor{grey83}{rgb}{0.83,0.83,0.83}
\definecolor{grey84}{rgb}{0.84,0.84,0.84}
\definecolor{grey85}{rgb}{0.85,0.85,0.85}
\definecolor{grey86}{rgb}{0.86,0.86,0.86}
\definecolor{grey87}{rgb}{0.87,0.87,0.87}
\definecolor{grey88}{rgb}{0.88,0.88,0.88}
\definecolor{grey89}{rgb}{0.89,0.89,0.89}
\definecolor{grey8}{rgb}{0.08,0.08,0.08}
\definecolor{grey90}{rgb}{0.90,0.90,0.90}
\definecolor{grey91}{rgb}{0.91,0.91,0.91}
\definecolor{grey92}{rgb}{0.92,0.92,0.92}
\definecolor{grey93}{rgb}{0.93,0.93,0.93}
\definecolor{grey94}{rgb}{0.94,0.94,0.94}
\definecolor{grey95}{rgb}{0.95,0.95,0.95}
\definecolor{grey96}{rgb}{0.96,0.96,0.96}
\definecolor{grey97}{rgb}{0.97,0.97,0.97}
\definecolor{grey98}{rgb}{0.98,0.98,0.98}
\definecolor{grey99}{rgb}{0.99,0.99,0.99}
\definecolor{grey9}{rgb}{0.09,0.09,0.09}
\definecolor{grey}{rgb}{0.75,0.75,0.75}
\definecolor{honeydew1}{rgb}{0.94,1.00,0.94}
\definecolor{honeydew2}{rgb}{0.88,0.93,0.88}
\definecolor{honeydew3}{rgb}{0.76,0.80,0.76}
\definecolor{honeydew4}{rgb}{0.51,0.55,0.51}
\definecolor{honeydew}{rgb}{0.94,1.00,0.94}
\definecolor{hotpink}{rgb}{1.00,0.41,0.71}
\definecolor{indianred}{rgb}{0.80,0.36,0.36}
\definecolor{ivory1}{rgb}{1.00,1.00,0.94}
\definecolor{ivory2}{rgb}{0.93,0.93,0.88}
\definecolor{ivory3}{rgb}{0.80,0.80,0.76}
\definecolor{ivory4}{rgb}{0.55,0.55,0.51}
\definecolor{ivory}{rgb}{1.00,1.00,0.94}
\definecolor{khaki1}{rgb}{1.00,0.96,0.56}
\definecolor{khaki2}{rgb}{0.93,0.90,0.52}
\definecolor{khaki3}{rgb}{0.80,0.78,0.45}
\definecolor{khaki4}{rgb}{0.55,0.53,0.31}
\definecolor{khaki}{rgb}{0.94,0.90,0.55}
\definecolor{lavenderblush}{rgb}{1.00,0.94,0.96}
\definecolor{lavender}{rgb}{0.90,0.90,0.98}
\definecolor{lawngreen}{rgb}{0.49,0.99,0.00}
\definecolor{lemonchiffon}{rgb}{1.00,0.98,0.80}
\definecolor{lightblue}{rgb}{0.68,0.85,0.90}
\definecolor{lightcoral}{rgb}{0.94,0.50,0.50}
\definecolor{lightcyan}{rgb}{0.88,1.00,1.00}
\definecolor{lightgoldenrod}{rgb}{0.93,0.87,0.51}
\definecolor{lightgoldenrod}{rgb}{0.98,0.98,0.82}
\definecolor{lightgray}{rgb}{0.83,0.83,0.83}
\definecolor{lightgreen}{rgb}{0.56,0.93,0.56}
\definecolor{lightgrey}{rgb}{0.83,0.83,0.83}
\definecolor{lightpink}{rgb}{1.00,0.71,0.76}
\definecolor{lightsalmon}{rgb}{1.00,0.63,0.48}
\definecolor{lightsea}{rgb}{0.13,0.70,0.67}
\definecolor{lightsky}{rgb}{0.53,0.81,0.98}
\definecolor{lightslate}{rgb}{0.47,0.53,0.60}
\definecolor{lightslate}{rgb}{0.47,0.53,0.60}
\definecolor{lightslate}{rgb}{0.52,0.44,1.00}
\definecolor{lightsteel}{rgb}{0.69,0.77,0.87}
\definecolor{lightyellow}{rgb}{1.00,1.00,0.88}
\definecolor{limegreen}{rgb}{0.20,0.80,0.20}
\definecolor{linen}{rgb}{0.98,0.94,0.90}
\definecolor{magenta1}{rgb}{1.00,0.00,1.00}
\definecolor{magenta2}{rgb}{0.93,0.00,0.93}
\definecolor{magenta3}{rgb}{0.80,0.00,0.80}
\definecolor{magenta4}{rgb}{0.55,0.00,0.55}
\definecolor{magenta}{rgb}{1.00,0.00,1.00}
\definecolor{maroon1}{rgb}{1.00,0.20,0.70}
\definecolor{maroon2}{rgb}{0.93,0.19,0.65}
\definecolor{maroon3}{rgb}{0.80,0.16,0.56}
\definecolor{maroon4}{rgb}{0.55,0.11,0.38}
\definecolor{maroon}{rgb}{0.69,0.19,0.38}
\definecolor{mediumaquamarine}{rgb}{0.40,0.80,0.67}
\definecolor{mediumblue}{rgb}{0.00,0.00,0.80}
\definecolor{mediumorchid}{rgb}{0.73,0.33,0.83}
\definecolor{mediumpurple}{rgb}{0.58,0.44,0.86}
\definecolor{mediumsea}{rgb}{0.24,0.70,0.44}
\definecolor{mediumslate}{rgb}{0.48,0.41,0.93}
\definecolor{mediumspring}{rgb}{0.00,0.98,0.60}
\definecolor{mediumturquoise}{rgb}{0.28,0.82,0.80}
\definecolor{mediumviolet}{rgb}{0.78,0.08,0.52}
\definecolor{midnightblue}{rgb}{0.10,0.10,0.44}
\definecolor{mintcream}{rgb}{0.96,1.00,0.98}
\definecolor{mistyrose}{rgb}{1.00,0.89,0.88}
\definecolor{moccasin}{rgb}{1.00,0.89,0.71}
\definecolor{navajowhite}{rgb}{1.00,0.87,0.68}
\definecolor{navyblue}{rgb}{0.00,0.00,0.50}
\definecolor{navy}{rgb}{0.00,0.00,0.50}
\definecolor{oldlace}{rgb}{0.99,0.96,0.90}
\definecolor{olivedrab}{rgb}{0.42,0.56,0.14}
\definecolor{orange1}{rgb}{1.00,0.65,0.00}
\definecolor{orange2}{rgb}{0.93,0.60,0.00}
\definecolor{orange3}{rgb}{0.80,0.52,0.00}
\definecolor{orange4}{rgb}{0.55,0.35,0.00}
\definecolor{orangered}{rgb}{1.00,0.27,0.00}
\definecolor{orange}{rgb}{1.00,0.65,0.00}
\definecolor{orchid1}{rgb}{1.00,0.51,0.98}
\definecolor{orchid2}{rgb}{0.93,0.48,0.91}
\definecolor{orchid3}{rgb}{0.80,0.41,0.79}
\definecolor{orchid4}{rgb}{0.55,0.28,0.54}
\definecolor{orchid}{rgb}{0.85,0.44,0.84}
\definecolor{palegoldenrod}{rgb}{0.93,0.91,0.67}
\definecolor{palegreen}{rgb}{0.60,0.98,0.60}
\definecolor{paleturquoise}{rgb}{0.69,0.93,0.93}
\definecolor{paleviolet}{rgb}{0.86,0.44,0.58}
\definecolor{papayawhip}{rgb}{1.00,0.94,0.84}
\definecolor{peachpuff}{rgb}{1.00,0.85,0.73}
\definecolor{peru}{rgb}{0.80,0.52,0.25}
\definecolor{pink1}{rgb}{1.00,0.71,0.77}
\definecolor{pink2}{rgb}{0.93,0.66,0.72}
\definecolor{pink3}{rgb}{0.80,0.57,0.62}
\definecolor{pink4}{rgb}{0.55,0.39,0.42}
\definecolor{pink}{rgb}{1.00,0.75,0.80}
\definecolor{plum1}{rgb}{1.00,0.73,1.00}
\definecolor{plum2}{rgb}{0.93,0.68,0.93}
\definecolor{plum3}{rgb}{0.80,0.59,0.80}
\definecolor{plum4}{rgb}{0.55,0.40,0.55}
\definecolor{plum}{rgb}{0.87,0.63,0.87}
\definecolor{powderblue}{rgb}{0.69,0.88,0.90}
\definecolor{purple1}{rgb}{0.61,0.19,1.00}
\definecolor{purple2}{rgb}{0.57,0.17,0.93}
\definecolor{purple3}{rgb}{0.49,0.15,0.80}
\definecolor{purple4}{rgb}{0.33,0.10,0.55}
\definecolor{purple}{rgb}{0.63,0.13,0.94}
\definecolor{red1}{rgb}{1.00,0.00,0.00}
\definecolor{red2}{rgb}{0.93,0.00,0.00}
\definecolor{red3}{rgb}{0.80,0.00,0.00}
\definecolor{red4}{rgb}{0.55,0.00,0.00}
\definecolor{red}{rgb}{1.00,0.00,0.00}
\definecolor{rosybrown}{rgb}{0.74,0.56,0.56}
\definecolor{royalblue}{rgb}{0.25,0.41,0.88}
\definecolor{saddlebrown}{rgb}{0.55,0.27,0.07}
\definecolor{salmon1}{rgb}{1.00,0.55,0.41}
\definecolor{salmon2}{rgb}{0.93,0.51,0.38}
\definecolor{salmon3}{rgb}{0.80,0.44,0.33}
\definecolor{salmon4}{rgb}{0.55,0.30,0.22}
\definecolor{salmon}{rgb}{0.98,0.50,0.45}
\definecolor{sandybrown}{rgb}{0.96,0.64,0.38}
\definecolor{seagreen}{rgb}{0.18,0.55,0.34}
\definecolor{seashell1}{rgb}{1.00,0.96,0.93}
\definecolor{seashell2}{rgb}{0.93,0.90,0.87}
\definecolor{seashell3}{rgb}{0.80,0.77,0.75}
\definecolor{seashell4}{rgb}{0.55,0.53,0.51}
\definecolor{seashell}{rgb}{1.00,0.96,0.93}
\definecolor{sienna1}{rgb}{1.00,0.51,0.28}
\definecolor{sienna2}{rgb}{0.93,0.47,0.26}
\definecolor{sienna3}{rgb}{0.80,0.41,0.22}
\definecolor{sienna4}{rgb}{0.55,0.28,0.15}
\definecolor{sienna}{rgb}{0.63,0.32,0.18}
\definecolor{skyblue}{rgb}{0.53,0.81,0.92}
\definecolor{slateblue}{rgb}{0.42,0.35,0.80}
\definecolor{slategray}{rgb}{0.44,0.50,0.56}
\definecolor{slategrey}{rgb}{0.44,0.50,0.56}
\definecolor{snow1}{rgb}{1.00,0.98,0.98}
\definecolor{snow2}{rgb}{0.93,0.91,0.91}
\definecolor{snow3}{rgb}{0.80,0.79,0.79}
\definecolor{snow4}{rgb}{0.55,0.54,0.54}
\definecolor{snow}{rgb}{1.00,0.98,0.98}
\definecolor{springgreen}{rgb}{0.00,1.00,0.50}
\definecolor{steelblue}{rgb}{0.27,0.51,0.71}
\definecolor{tan1}{rgb}{1.00,0.65,0.31}
\definecolor{tan2}{rgb}{0.93,0.60,0.29}
\definecolor{tan3}{rgb}{0.80,0.52,0.25}
\definecolor{tan4}{rgb}{0.55,0.35,0.17}
\definecolor{tan}{rgb}{0.82,0.71,0.55}
\definecolor{thistle1}{rgb}{1.00,0.88,1.00}
\definecolor{thistle2}{rgb}{0.93,0.82,0.93}
\definecolor{thistle3}{rgb}{0.80,0.71,0.80}
\definecolor{thistle4}{rgb}{0.55,0.48,0.55}
\definecolor{thistle}{rgb}{0.85,0.75,0.85}
\definecolor{tomato1}{rgb}{1.00,0.39,0.28}
\definecolor{tomato2}{rgb}{0.93,0.36,0.26}
\definecolor{tomato3}{rgb}{0.80,0.31,0.22}
\definecolor{tomato4}{rgb}{0.55,0.21,0.15}
\definecolor{tomato}{rgb}{1.00,0.39,0.28}
\definecolor{turquoise1}{rgb}{0.00,0.96,1.00}
\definecolor{turquoise2}{rgb}{0.00,0.90,0.93}
\definecolor{turquoise3}{rgb}{0.00,0.77,0.80}
\definecolor{turquoise4}{rgb}{0.00,0.53,0.55}
\definecolor{turquoise}{rgb}{0.25,0.88,0.82}
\definecolor{violetred}{rgb}{0.82,0.13,0.56}
\definecolor{violet}{rgb}{0.93,0.51,0.93}
\definecolor{wheat1}{rgb}{1.00,0.91,0.73}
\definecolor{wheat2}{rgb}{0.93,0.85,0.68}
\definecolor{wheat3}{rgb}{0.80,0.73,0.59}
\definecolor{wheat4}{rgb}{0.55,0.49,0.40}
\definecolor{wheat}{rgb}{0.96,0.87,0.70}
\definecolor{whitesmoke}{rgb}{0.96,0.96,0.96}
\definecolor{white}{rgb}{1.00,1.00,1.00}
\definecolor{yellow1}{rgb}{1.00,1.00,0.00}
\definecolor{yellow2}{rgb}{0.93,0.93,0.00}
\definecolor{yellow3}{rgb}{0.80,0.80,0.00}
\definecolor{yellow4}{rgb}{0.55,0.55,0.00}
\definecolor{yellowgreen}{rgb}{0.60,0.80,0.20}
\definecolor{yellow}{rgb}{1.00,1.00,0.00}

\begin{document}
\bibliographystyle{JHEP}	

\title{Measuring the X-ray Background in the Reionization Era with First Generation 21 cm Experiments}
\author[1]{Pierre Christian%
\note{pchristian@cfa.harvard.edu}}
\affiliation{Astronomy Department, Harvard University,\\
60 Garden Street, Cambridge, MA}
 \author{and Abraham Loeb}
\notoc 

\abstract{The X-ray background during the epoch of reionization is currently poorly constrained. We demonstrate that it is possible to use first generation 21 cm experiments to calibrate it. Using the semi-numerical simulation, 21cmFAST, we calculate the dependence of the 21 cm power spectrum on the X-ray background flux. Comparing the signal to the sensitivity of the Murchison Widefield Array (MWA) we find that in the redshift interval $z=$$8$-$14$ the 21 cm signal is detectable for certain values of the X-ray background. We show that there is no degeneracy between the X-ray production efficiency and the Ly$\alpha$ production efficiency and that the degeneracy with the ionization fraction of the intergalactic medium can be broken.}

\begin{onecolumn}

\date{\today}
\maketitle

\end{onecolumn}

\section{Introduction}

The dominant heating process of the intergalactic medium (IGM) during the epoch of reionization (EoR) is mediated by X-rays \citep{PritchardReview}. Through photoionization of HI and HeI, X-ray photons generate high energy electrons which then deposit their energy into increasing the gas temperature. In principle, ultraviolet photons, which constitute the bulk of ionizing radiation during the EoR, could contribute to this photo-heating process. However, their mean free path is much smaller than the X-ray photons', thereby constraining their photo-heating to the ionized regions around their sources \citep{Dijkstra}.

There are multiple hypotheses concerning the sources of the X-ray photons, including active galactic nuclei \citep{AGN}, miniquasars \citep{miniquasar}, and high mass X-ray binaries \citep{mineo}. In principle, if we knew both the number density and the X-ray emissivity of these sources during the EoR, we could have inferred the EoR X-ray background. This approach however embodies many uncertainties due to unaccounted X-ray sources \citep{Dijkstra}.

Recently, there has been a growing interest in using the cosmologically redshifted 21 cm signal from neutral hydrogen to probe the reionization era. The potential of the redshifted 21 cm emission to be an emissary of the EoR motivated the construction of several radio experiments such as the Murchison Widefield Array (MWA)\footnote{http://www.MWAtelescope.org/},  the Precision Array for Probing the Epoch of Reionization (PAPER)\footnote{http://eor.berkeley.edu/}, and the LOw Frequency ARray (LOFAR)\footnote{http://www.lofar.org/}; all of which just started their operation.

During the EoR, the intensity of the redshifted 21 cm signal is primarily modulated by a combination of density, ionization fraction, and spin temperature fluctuations \citep{Pritchard_Furlanetto_2007}. The sensitivity of the spin temperature to the gas temperature makes it natural to consider employing the redshifted 21 cm signal to measure the EoR X-ray background (XRB) intensity. In fact, a recent upper limit published by the PAPER collaboration had already ruled out an EoR heating scenario with no X-ray heating \citep{PAPERlimit}. 

In this work we examine whether the sensitivities of first generation radio arrays are sufficient to provide constraints to the EoR X-ray background intensity. Although we focus our work on the 128 antennae of the MWA, our results can be generalized to other first generation observatories. Throughout this paper, we will adopt the $\Lambda$CDM cosmological parameters: $\Omega_M=0.32$ , $\Omega_b=0.05$, $\Omega_\Lambda = 0.68$, and $h=0.67$ \citep{Planck}.

\section{The 21 cm Signal}
The 21 cm signal results from the spin flip transition of the electron in neutral hydrogen. The intensity of this signal is quantified in the offset of the 21 cm brightness temperature, $T_b$, from the Cosmic Microwave Background (CMB) temperature, $T_\gamma$. This offset at a given frequency $\nu$ along a line of sight at a redshift $z$ is given by \citep{LoebBook}:
\begin{equation}
\delta T_b(\nu) = \frac{T_S - T_\gamma } {1+z} (1 - e^{-\tau_{\nu_0}})  \approx  
27 x_{HI} (1 + \delta) \left(  \frac{H}  {dv_r /dr + H} \right) \left(  1 - \frac{T_\gamma} {T_S} \right) 
\times 0.3153 \left(1+z  \right)^{1/2} \;  \rm mK,
\end{equation}
where $z$ is the signal's redshift, $T_S$ the spin temperature of the gas, $\tau_{\nu_0}$ is the optical depth at the 21 cm line frequency, $x_{HI}=1 - x_i$ is the neutral fraction (where $x_i$ is the ionized fraction), $\delta (\vec{x}, z) \equiv \rho/ \bar{\rho} -1$ the fractional overdensity, $H(z)$ the Hubble parameter, and $dv_r/dr$ the comoving velocity gradient along the line of sight. Note that the intensity of the 21 cm signal is proportional to $T_S - T_\gamma$. Higher value of $|T_S - T_\gamma|$ will result in a stronger 21 cm signal. Regions where $T_S < T_\gamma$ will appear in absorption, while regions with $T_S > T_\gamma$ will appear in emission.

Three effects regulate the spin temperature $T_S$ \citep{LoebBook}: collisions between hydrogen atoms, as well as with free protons and electrons; scattering of CMB photons; and the Wouthuysen-Field effect, which describes the mixing of the hyperfine states due to absorption and re-emission of Ly$\alpha$ photons \citep{Wouthuysen} \citep{Field}. The combination of these effects results in $T_S$ that is given by the equation \citep{Field}:
\begin{equation}
\frac{1}{T_s} = \frac{T_\gamma^{-1} + x_\alpha T_c^{-1} + x_c T_K^{-1}} {1 + x_\alpha + x_c} \; ,
\end{equation}
where $x_c$ and $x_\alpha$ are the coupling coefficients describing collisions and Ly{$\alpha$} scattering, respectively, $T_c$ is the color temperature of the Ly$\alpha$ photons, and $T_K$ is the kinetic temperature of the gas. 

During the reionization era, the universe is not sufficiently dense for collisional coupling to be efficient. However, multiple scattering couple the Ly{$\alpha$} color temperature to the kinetic temperature so that $T_c \sim T_K$ \citep{PritchardReview}. In this paper, the most important aspect of $T_S$ is this dependency on $T_K$. X-ray photons are the dominant source of heating during the EoR, and as such the X-ray background during the EoR is strongly coupled to the kinetic temperature, $T_K$.

\section{21cmFAST Simulation}

We use the output of the 21cmFAST code, a semi-numerical simulation of the redshifted 21 cm which takes into account the evolution of $\delta T_b$ as prescribed in equation (1) \citep{21cmFAST}. Our simulation volume is a 400$\times$400$\times$400 comoving Mpc box discretized to 800 cells per axis. Our simulations ran from $z=35$ to $z=7$.

In order to explore the X-ray background efficiency, 21cmFAST parameterizes the X-ray background efficiency by the two parameters  $\zeta_x$, the number of X-ray photons emitted per solar mass in stars and $f_{\star}$, the star formation efficiency. The effects of these two parameters are multiplicative, so the effective X-ray background efficiency is proportional to the product $\zeta_x f_{\star}$, the number of X-ray photons emitted per total baryonic mass in collapsed halos.

We ran the simulation multiple times with different values of $\zeta_x$. In practice, our results are insensitive to whether the change in X-ray background is due to a change in $\zeta_x$ or a change in $f_{\star}$. Changing $\zeta_x$ alone while keeping $f_{\star}$ constant is equivalent to altering their product. The fiducial value often assumed for this parameter is $\zeta_x f_{\star} = 10^{56}$ photons per solar mass in stars. This corresponds to $\sim 1$ X-ray photon per stellar baryon with $f_{\star} = 10\%$, the value inferred at $z=0$.

\subsection{Generalization of X-ray Sources}

We can translate $\zeta_x f_{\star}$ native to 21cmFAST to more intuitive intensity units by the following change of variable:
\begin{equation}
I_{X} = \frac{c}{4 \pi} \times \frac{\Omega_{b}}{\Omega_{M}} \times \zeta_x f_{\star} \times \rho_{coll} \times \langle h \nu \rangle ,
\end{equation}
where $I_{X}$ is the X-ray intensity, $\rho_{coll}$ the global mass density in collapsed halos, and $\langle h \nu \rangle$ the average photon energy (taken to be $\sim 1 \; \rm  keV$). Note that in general this relation depends on redshift.

We calculate $\rho_{coll}$ from the Sheth-Tormen mass function of collapsed dark matter halos \citep{2002MNRAS.329...61S}:
\begin{equation}
n_{ST}(M) = A' \sqrt{\frac{2a'}{\pi}} \frac{\rho_m} {M}  \frac{-d (ln \; \sigma)}{dM} \nu_c \left[ 1 + \frac{1}{(a' \nu_c^2)^{q'}} \right]  
\times \exp\left( \frac{-a' \nu_c^2}{2} \right ) \; ,
\end{equation}
with $a'=0.707$, $q'=0.3$, and $A'=0.322$, fitted to model ellipsoidal collapse. $\rho_{coll}$ is the integration of $n_{ST}$ as a function of mass:
\begin{equation}
\rho_{coll} = \int_{M{\min}}^\infty M n_{sT} (M) dM \; ,
\end{equation}
where we adopt a minimum galaxy halo mass, $M_{min}$, corresponding to the threshold for atomic hydrogen cooling at a virial temperature of $\sim 10^4$ K.

\subsection{Conversion between $\zeta_X$ and $f_X$}
It is often more convenient to express the X-ray background in terms of $f_X$, the parameter connecting the X-ray luminosity, $L_X$, with the star formation rate, SFR \citep{Dijkstra}:

\begin{equation}
L_X = 1.4 \times 10^{39} f_X f_{\star} \left( \frac{\rm SFR}{M_{\odot} \; \rm yr^{-1}} \right) \rm erg \; s^{-1} \; ,
\end{equation}
instead of $\zeta_X$. We can relate $\zeta_X$ and $f_X$ by normalizing $f_X=1$ to the $L_X$-SFR relation of Dijkstra et al. (2012) \cite{Dijkstra}: $L_X = 1.4 \times 10^{39} \left( \frac{SFR} {M_{\odot} \; \rm yr^{-1}} \right) \rm erg \; s^{-1}$ when $f_X f_{\star}=1$. The resulting conversion is: 
\begin{equation}
f_X =   2.14 \left( \frac{\zeta_x} {10^{57} \; \rm M_{\odot}^{-1}} \right) \; .
\end{equation}
We will use $f_X$ in place of $\zeta_x$ throughout the rest of the paper.

\section{Noise}
The 21 cm power spectrum is inherently three dimensional. In Fourier space, $u$ and $v$ correspond to the two angular dimensions while $f$ is the frequency of the observation (or redshift) providing the line-of-sight spatial dimension. For each $(u, v, f)$ voxel, there is a corresponding thermal noise that we must calculate to figure out the sensitivity of the observation. 

We start with the radiometer equation for the thermal noise on the visibility measurement \citep{Moran}:
\begin{equation}
V_{rms} (u, v, f) = \frac{c^2 T_{sys}} {f^2 A_e \sqrt{\Delta f \tau}} \; ,
\end{equation}
where $T_{sys}$ is the system temperature, assumed to be uniform across the observational bandwidth, $A_e$ the effective area per antenna, $\Delta f$ the channel width, and $\tau$ is the total integration time for the specific $(u, v, f)$ voxel. It is important to note that due to the redundancy of the baselines, $\tau$ can be very different from the actual observation time of the array. 
  
At this point, it is useful to bring our calculation to the familiar cosmological wavenumber space, $\vec{k}$. Following Beardsley et al. (2013) \cite{Beardsley}, the $(u, v)$ angular frequencies can be mapped directly to $\vec{k}_{\perp}$, the
wavenumber perpendicular to the line-of-sight, by the mapping
\begin{equation}
k_{\perp} = \frac{2 \pi u}{D} \; ,
\end{equation}
where $D(z)$ is the comoving distance to a redshift $z$. To obtain a similar mapping for the line-of-sight direction, we map the observed frequency to redshift by using the rest frame frequency of the 21 cm line. Then we calculate the comoving distance to this redshift before performing a Fourier transform to bring it to wavenumber space. To obtain the noise on the power spectrum, we square equation (1) to obtain, as in Beardsley et al. (2013) \cite{Beardsley}:
\begin{equation}
C(\vec{k}) = T_{sys}^2 \left(\frac{D^2 \lambda^2} {A_e} \right) \left( \frac{\Delta D } {B} \right) \frac{1}{\tau(\vec{k})} \; .
\end{equation}
As stated previously, $\tau(\vec{k})$ is different from the actual observation time and is a function of the wavenumber $\vec{k}$. In particular, the functional form of $\tau(\vec{k})$ depends on the antenna positions. This work uses $\tau(\vec{k})$ as calculated by Beardsley et al. (2013) \cite{Beardsley}, which is specific to the 128 antennae of the MWA. Note, however, that $\tau$ still scales linearly with observation time, and as such, our noise goes down as $1/t$. Our values of $T_{sys}$, $A_e$, $\Delta D$, and $B$ are those listed in Table 1 of Beardsley et al. (2013)  \cite{Beardsley}. Averaging this noise in the three dimensional $\vec{k}$ space in spherical bins in the manner of McQuinn et al. (2006) \cite{McQuinn} gives us our noise in the one dimensional $|\vec{k}|$ space. 

\section{Results}

\begin{figure*}
  \centering
  \subfloat[$f_X f_{\star}=10^{-1}$]{\label{x57}\includegraphics[scale=0.5]{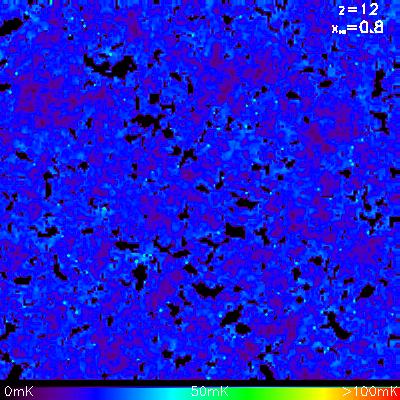}} \;
  \subfloat[$f_X f_{\star}=10^{-2}$]{\label{x56}\includegraphics[scale=0.5]{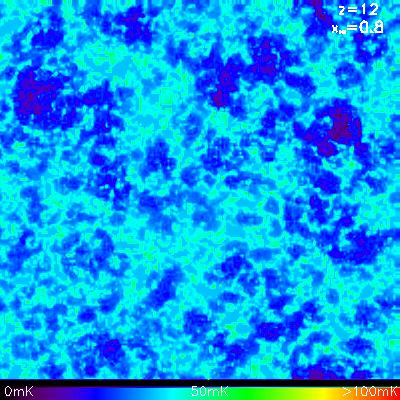}}   \\
  \subfloat[$f_X f_{\star}=10^{-3}$]{\label{x55}\includegraphics[scale=0.5]{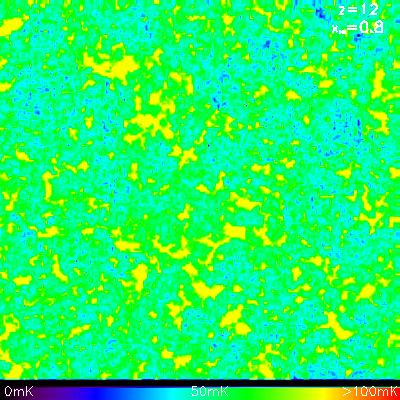}} \;
   \subfloat[$f_X f_{\star}=10^{-4}$]{\label{x54}\includegraphics[scale=0.5]{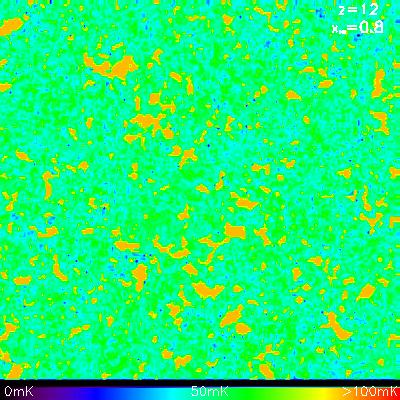}} 
  \caption{Slices of the 21 cm signal at $z=12$ for different values of $f_X f_{\star}$ with fiducial $f_{\alpha}=1$ and $\zeta_{ion}=31.5$. Note the monotonic increase in overall power as we decrease $f_X f_{\star}$.}
  \label{realization}
\end{figure*}

Slices of the 21 cm signal at $z=12$ from simulation runs with $f_X f_{\star}= 10^{-1}, \; 10^{-2}, \; 10^{-3}, \; 10^{-4} \; $ are shown in Figure \ref{realization}. Note that at these conditions, decreasing $f_X f_{\star}$ monotonically increases the overall 21 cm signal. The reason for this can be seen in the global evolution of the 21 cm signal. Decreasing the X-ray background delays gas heating, in effect moving the absorption trough to a lower redshift. Within the absorption trough, different regions can have very large differences in $T_K$, resulting in a large 21 cm intensity.

\begin{figure}
  \centering
  \subfloat[]{\label{fxfstar_zevolve}\includegraphics[scale=0.4]{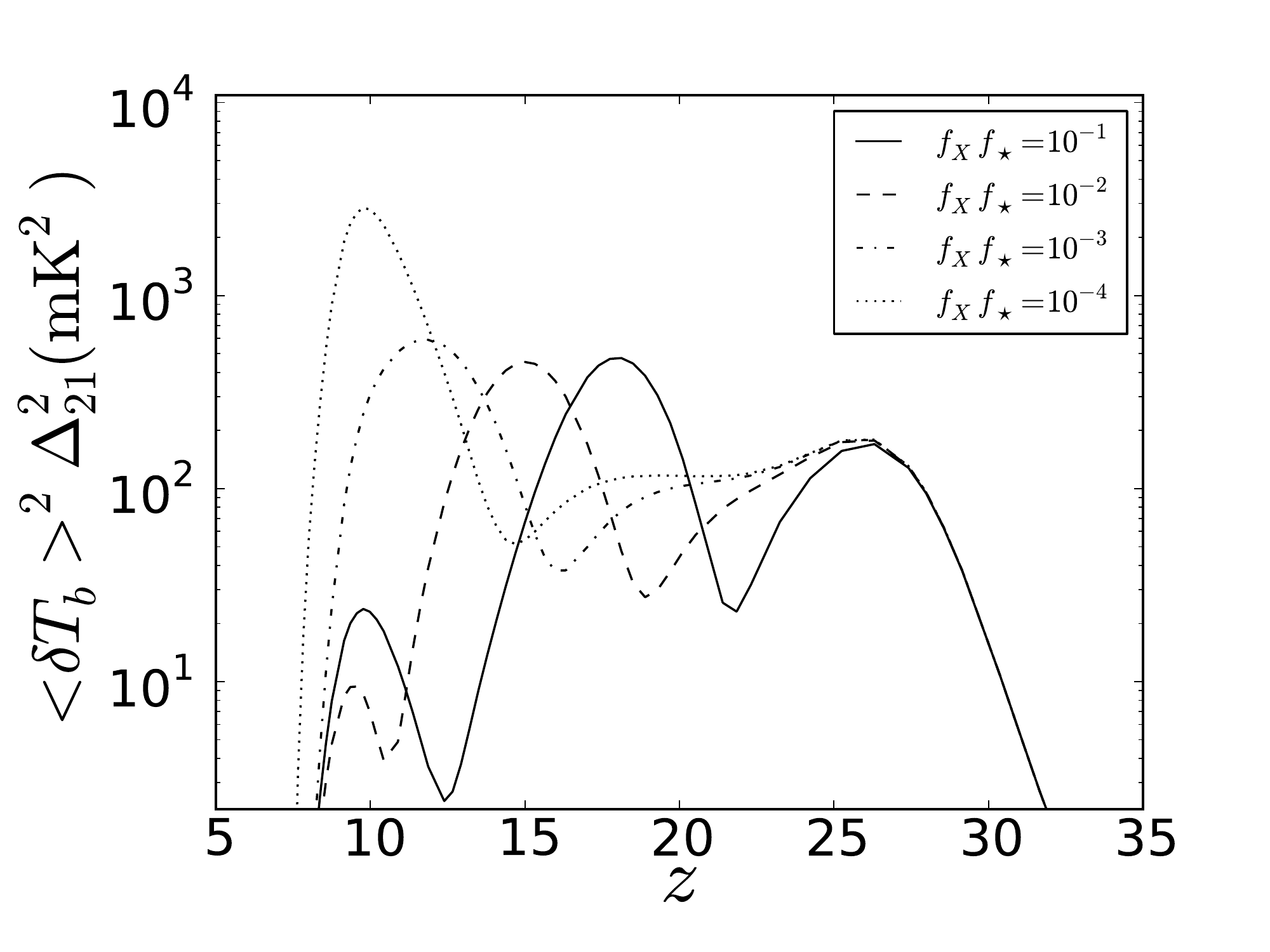}} 
  \subfloat[]{\label{zion_evolve}\includegraphics[scale=0.4]{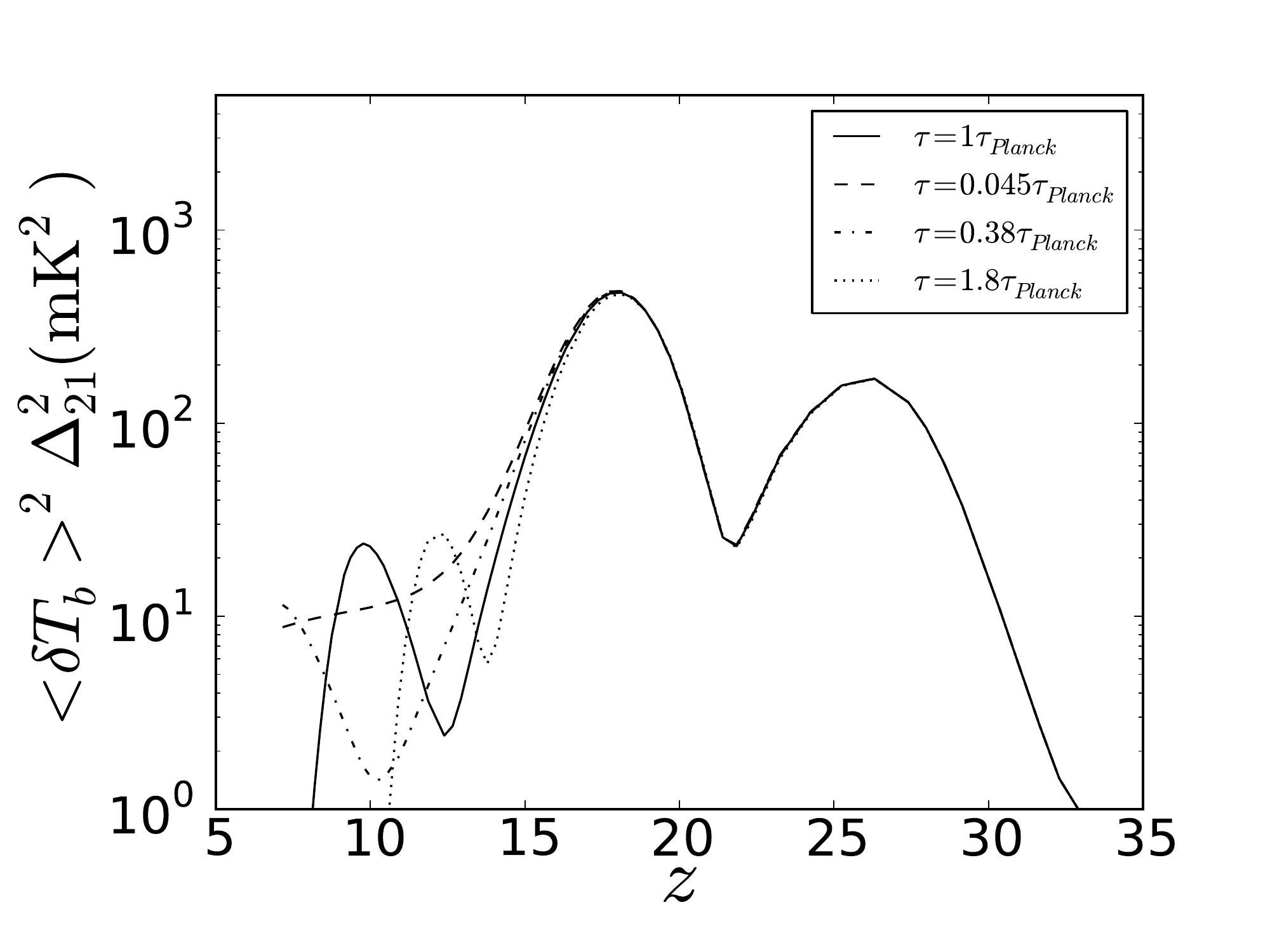}} 	
  \caption{(a) Evolution of the 21 cm power spectrum at $k=0.1 \; \rm Mpc^{-1}$ as a function of redshift. The three peaks prominently shown by the fiducial model $f_X f_{\star}= 10^{-1}$ correspond to the epochs of Ly$\alpha$ pumping, X-ray heating, and reionization (from highest to lowest $z$). As the X-ray efficiency is decreased, the X-ray peak shifts towards lower redshifts, thereby putting it in a regime detectable by first generation observatories. The largest 21 cm signal possible is achieved at low X-ray efficiency where the X-ray peak merges with the reionization peak (e.g.  $f_X f_{\star}= 10^{-3} \; \& \; 10^{-4}$). (b) Same as (a), but modulating the ionization efficiency, $\zeta_{ion}$, while keeping $f_X f_{\star} = 10^{-1}$. Each model's optical depth is calculated using equation (10), and all optical depths are measured at $z=7$. The degeneracy between the ionization and X-ray efficiency can be broken by noting the qualitative difference between the curves of (a) and (b): changing $\zeta_{ion}$ will not change the location of the X-ray peak. One can use either the X-ray peak's redshift or the value of the power spectrum at the peak to uniquely determine the X-ray background $f_X f_{\star}$.}
  \label{zevolve}
\end{figure}

A clear visualization of this argument is presented in panel (a) of Figure \ref{zevolve}, which shows the 21 cm power spectrum at $k=0.1 \; \rm Mpc^{-1}$. As discussed in Mesinger et al. (2013) \cite{xray_mesinger}, there are three peaks in the signal corresponding to epochs of Ly$\alpha$ pumping, X-ray heating, and reionization (from highest to lowest $z$). As one decreases the X-ray efficiency, the X-ray peak is shifted towards low redshifts. At low enough X-ray efficiency (e.g. $f_X f_{\star} \le 10^{-3}$), the X-ray peak enters a redshift regime that is uncontaminated by the strong foreground \citep{FourFundamental} and thus is accessible by first generation experiments. 

\begin{figure}
  \centering
  \subfloat[$z=14$]{\label{z14}\includegraphics[scale=0.4]{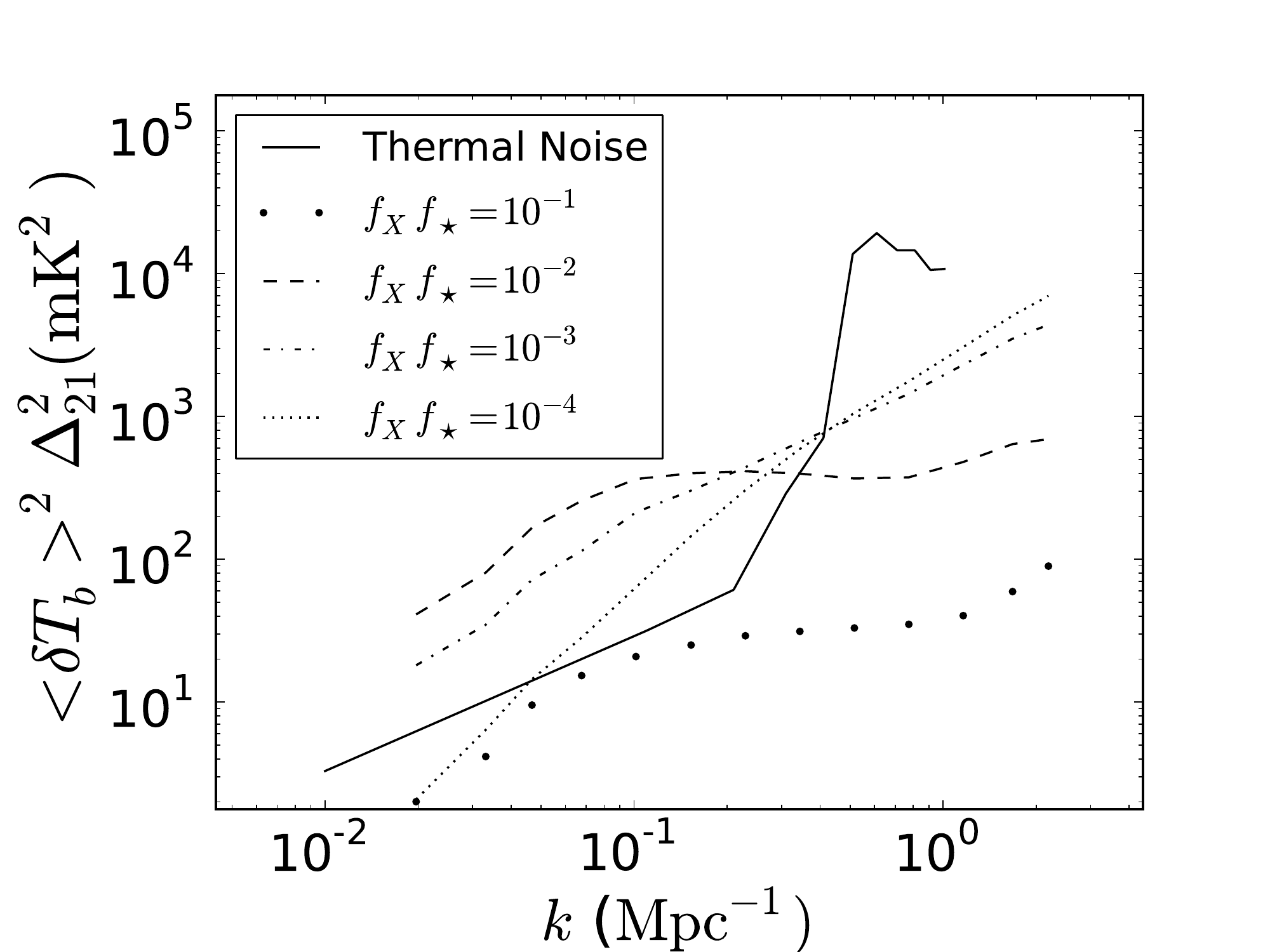}} 
  \subfloat[$z=12$]{\label{z12}\includegraphics[scale=0.4]{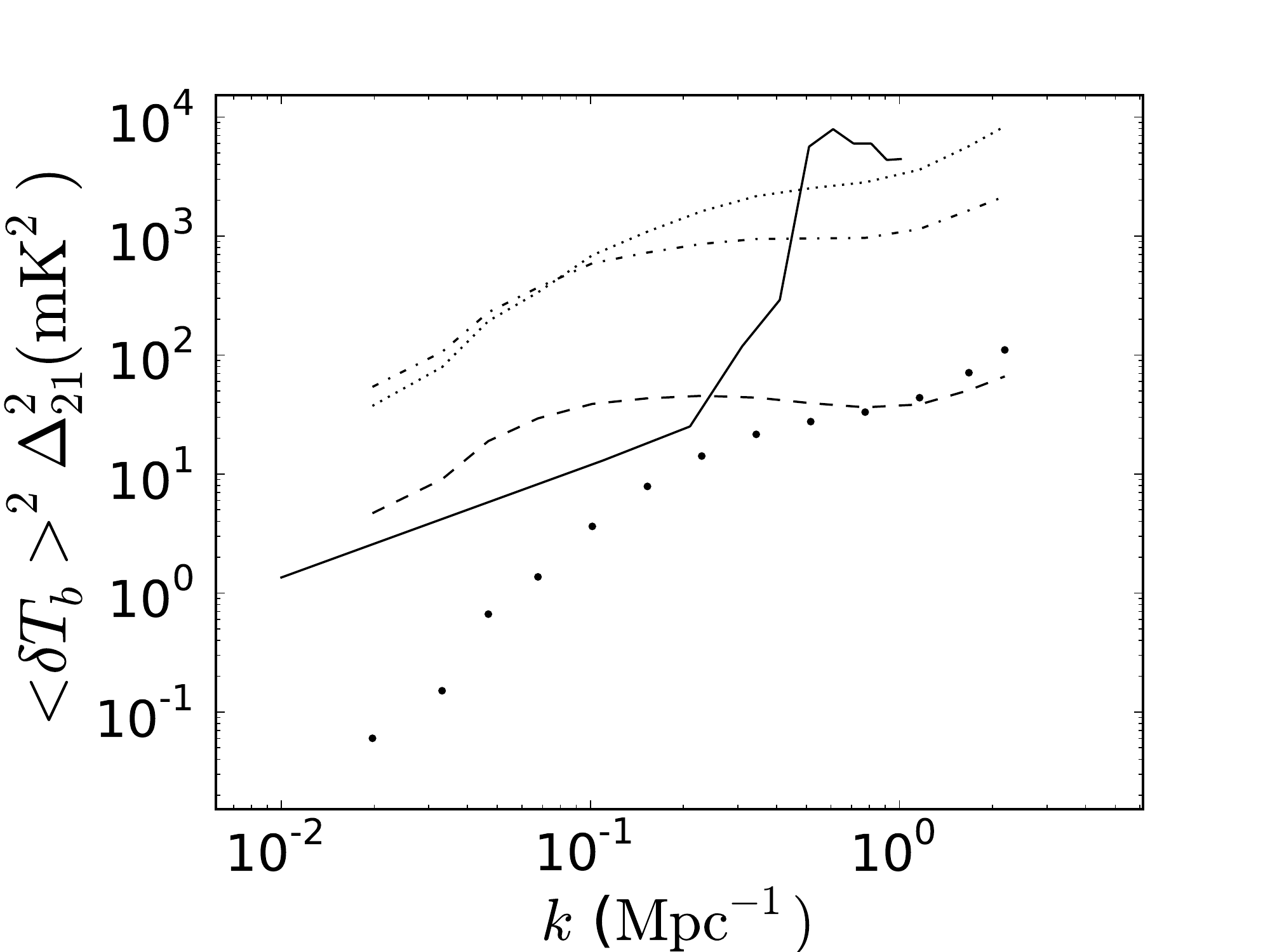}}   \\
  \subfloat[$z=10$]{\label{z10}\includegraphics[scale=0.4]{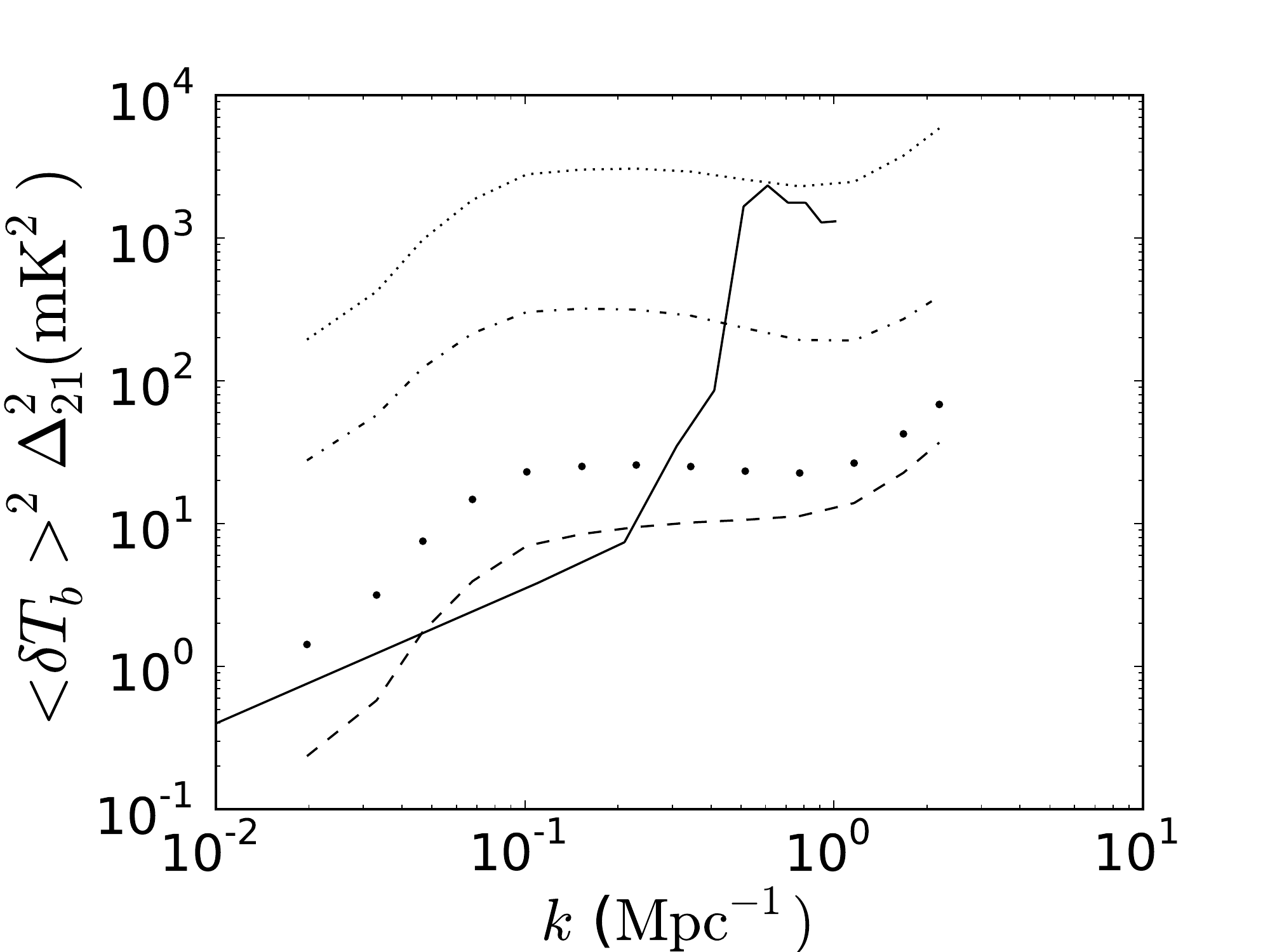}} 
   \subfloat[$z=8$]{\label{z8}\includegraphics[scale=0.4]{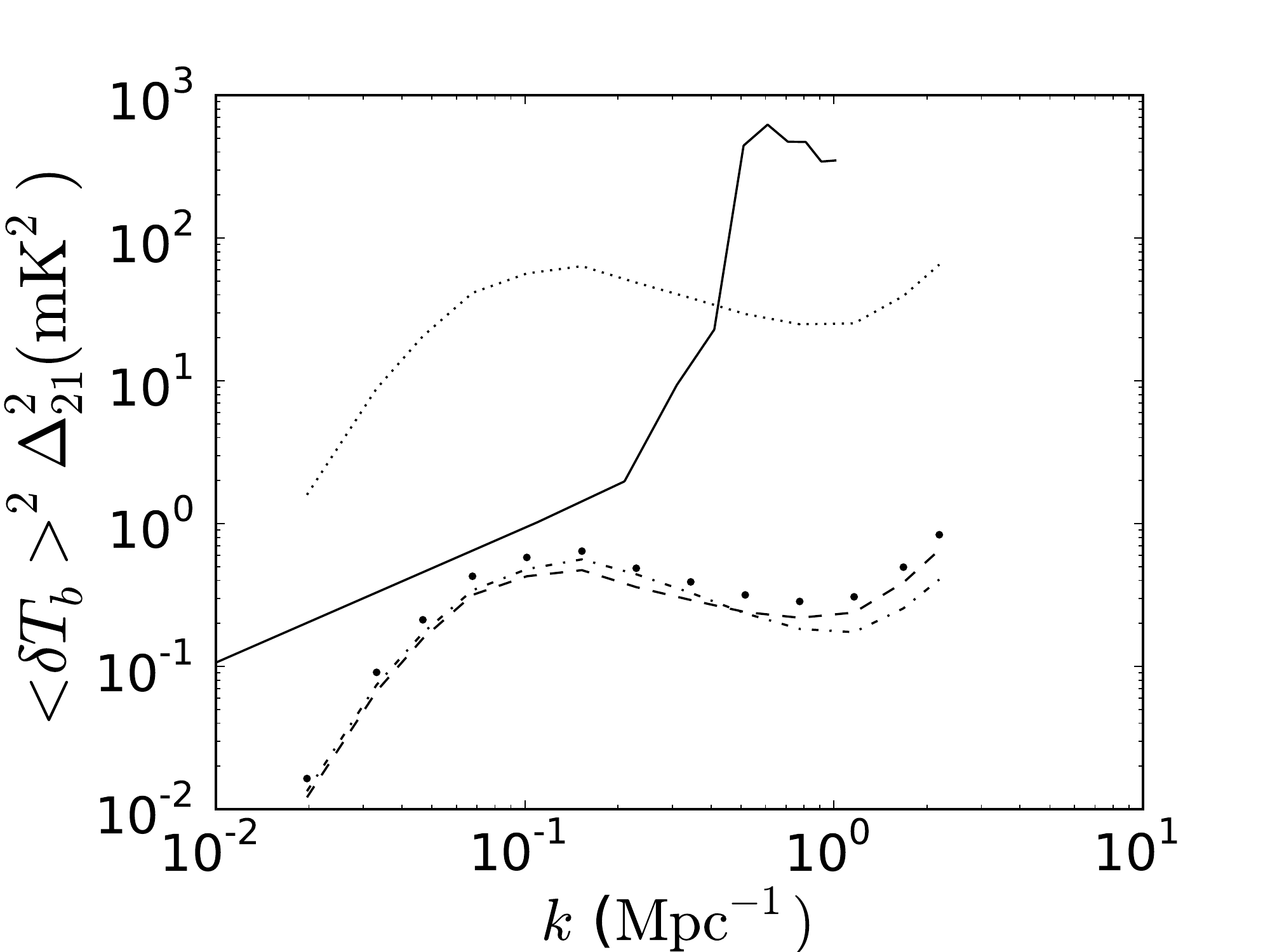}} 
  \caption{The thermal noise for 1000 hours of integration time with the MWA (solid line) and the 21 cm signal given a variety of X-ray efficiencies: $f_X f_{\star}=10^{-1},\;  10^{-2}, \; 10^{-3}, \; 10^{-4} $ for the points, dashed, dotted-dashed, and dotted lines respectively.}
  \label{result}
\end{figure}

The 21 cm power spectra and thermal noise for 1000 hours of MWA observation is shown in Figure \ref{result}. The sensitivity of this first generation array is high enough to place constraints on the X-ray background at redshift $z=8-14$. For example, at $k = 0.1 \; \rm Mpc^{-1}$, a non detection at $z=8$ will rule out reionization models with an X-ray background corresponding to $f_X f_{\star}=10^{-4} $ (dotted curve). A detection could be checked against the same set of simulations to ascertain the value of $f_X f_{\star}$ at the signal's redshift. As such, the 21 cm power spectrum could be used to measure the cosmic X-ray background at a given redshift during the EoR - making it an \textit{X-rayometer}.

Note that one does not need to measure the amplitude of the 21 cm power spectrum in order to measure the X-ray background. Figure \ref{zevolve} implies that the location of the X-ray peak is uniquely determined by the X-ray efficiency. The X-ray background is measurable from the redshift of this peak. Of course, measuring the actual value of the 21 cm power spectrum will allow us a more precise determination of the X-ray background, and the best way to locate the X-ray peak in the first place would involve measuring the 21 cm power spectrum around the peak. Still, lower limits on the X-ray background can be deduced by the absence of the X-ray peak below a certain redshift.

\subsection{Detectability of the 21 cm Signal}

The best prospects for detecting the 21 cm signal are represented by the peaks displayed in Figure \ref{zevolve}. The large X-ray peak for the fiducial model is at a high redshift, $z\sim 18$. At such high redshift, the foreground is large, preventing observations by first generation 21 cm observatories. However, lower X-ray backgrounds can delay the heating enough that the X-ray peak is moved to lower redshift. At $f_X f_{\star} = 10^{-2}$, for example, the X-ray peak is located at $z\sim 12$, a redshift accessible by first generation arrays.

The largest signal possible is located around $z=10-12$ at very low X-ray efficiencies ($f_X f_{\star}=10^{-3} \; \& \; 10^{-4}$), where the X-ray peak merges with the reionization peak.

\begin{figure}
  \centering
  \subfloat[]{\label{undiv}\includegraphics[scale=0.4]{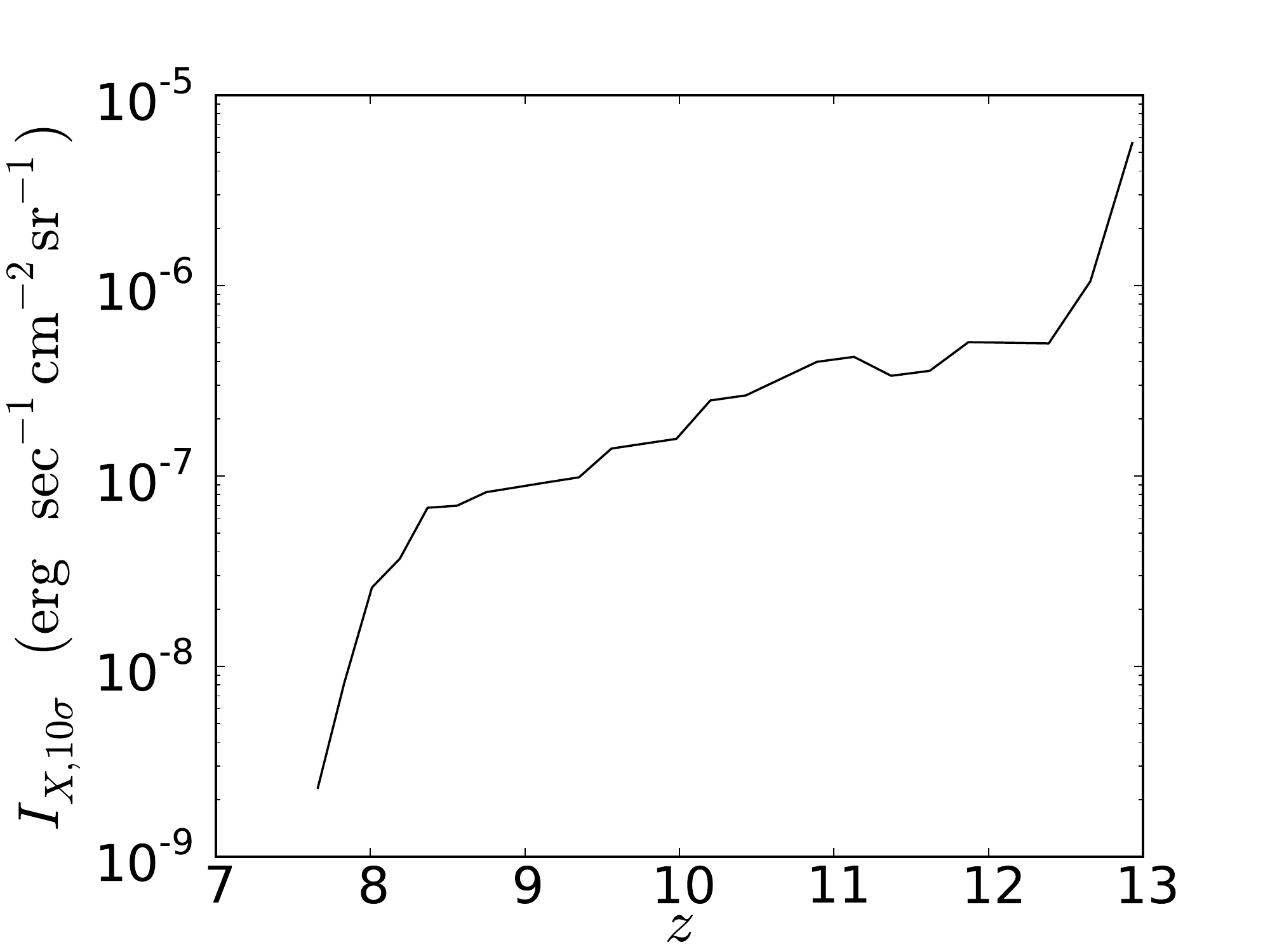}} 
  \subfloat[]{\label{div}\includegraphics[scale=0.4]{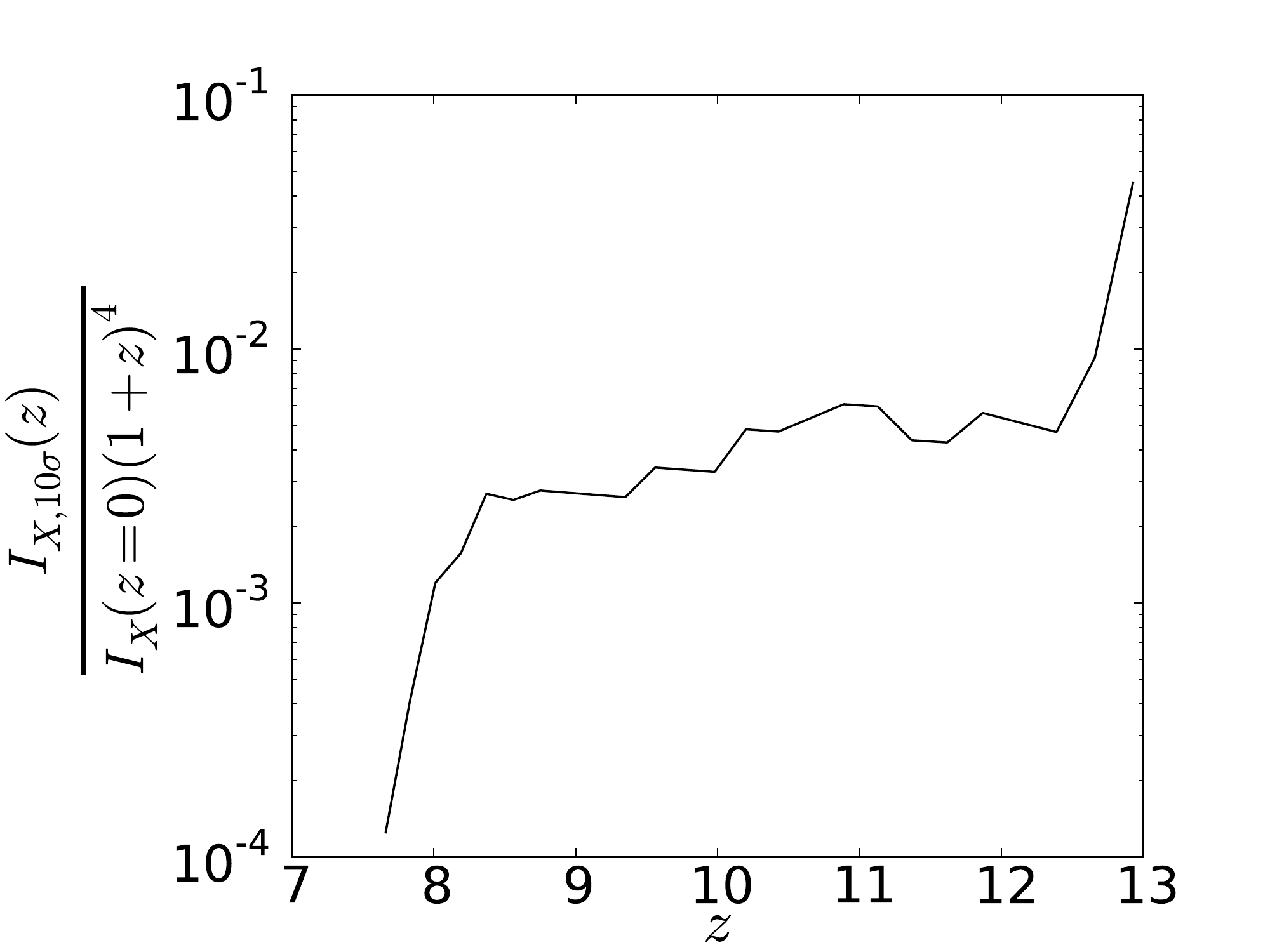}}   \\
  \caption{(a) The maximum X-ray background intensity required in order for the 21 cm signal at $k=0.1 \; \rm Mpc^{-1}$ be detectable at $10 \sigma$ by 1000 hours of MWA observations. We have used equation (5) to convert $f_X f_{\star}$ to physical units, $I_X$, the X-ray background intensity measured by a comoving observer at that redshift. (b) Same as (a), calculated as a fraction of the unresolved, diffuse soft X-ray background at $z=0$: $I_{X}(z=0) = 1 \pm 0.2 \times 10^{-12} \; \rm erg \; s^{-1} \; \rm cm^{-2} \; \rm deg^{-2}$. In comparing the X-ray background during the EoR and at $z=0$, we multiply the EoR values by $(1+z)^{-4}$ to account for the redshift dimming factor.}
  \label{detectability}
\end{figure}

Panel (a) of Figure \ref{detectability} presents the maximum X-ray background intensity allowed to make the 21 cm signal at $k=0.1 \; \rm Mpc^{-1}$ detectable at $10 \sigma$ after 1000 hours of MWA observations. Note that in using equation (5) to convert $f_X f_{\star}$ to physical units, we have placed our observer at the signal's redshift (i.e. $I_X$ measures the X-ray background intensity observed by a comoving observer at that redshift). In panel (b) of Figure \ref{detectability}, we calculated $I_X$ as a fraction of the observed unresolved soft X-ray background at $z=0$: $I_{X}(z=0) = 1 \pm 0.2 \times 10^{-12} \; \rm erg \; s^{-1} \; \rm cm^{-2} \; \rm deg^{-2}$ \citep{HandM_xray_z0}. We multiplied the EoR X-ray background by $(1+z)^{-4}$ to account for the redshift effect due to the expansion of the universe. 

Note that the value used for $I_{X}(z=0)$ is measured in the band between $0.65-1$ keV. We need to bear in mind that EoR X-ray photons will redshift away from these bands at $z=0$. Given a specific source spectrum one may correct the conversion between the emitted and observed X-ray intensity through a redshift factor for the corresponding energy bands. For active galactic nuclei at $z \sim 10$, this correction amounts to a modest reduction by a factor of $\sim 1.6$ \citep{AGN_spectra} in the vertical axis of Figure \ref{div}. For high-mass X-ray binaries, the correction is even weaker \citep{Dijkstra}. 

The maximum value of the X-ray background required for a $10 \sigma$ detection is much smaller than that allowed by the observed unresolved soft X-ray background. We have not taken into account, however, the possibility that most of the unresolved component of the $z=0$ soft X-ray background is produced by local sources \citep{Cappelluti}. 

\subsection{Monotonicity}

The aforementioned study is only possible in certain redshift regimes. The fourth panel of Figure \ref{result} shows the power spectra at $z=8$. At this redshift, the magnitude of the 21 cm power spectrum is insensitive to the X-ray background except when the efficiency is extremely low. For most of our choices of $f_X f_{\star}$ at $z \sim8$, the heating had saturated, thus rendering the power spectra insensitive to changes in $f_X f_{\star}$. Note that at very low X-ray backgrounds (e.g. $f_X f_{\star}= 10^{-4}$), the heating had not saturated at z=8 and a large signal amplitude is generated. 

\begin{figure}
\centering
\includegraphics[scale=0.43]{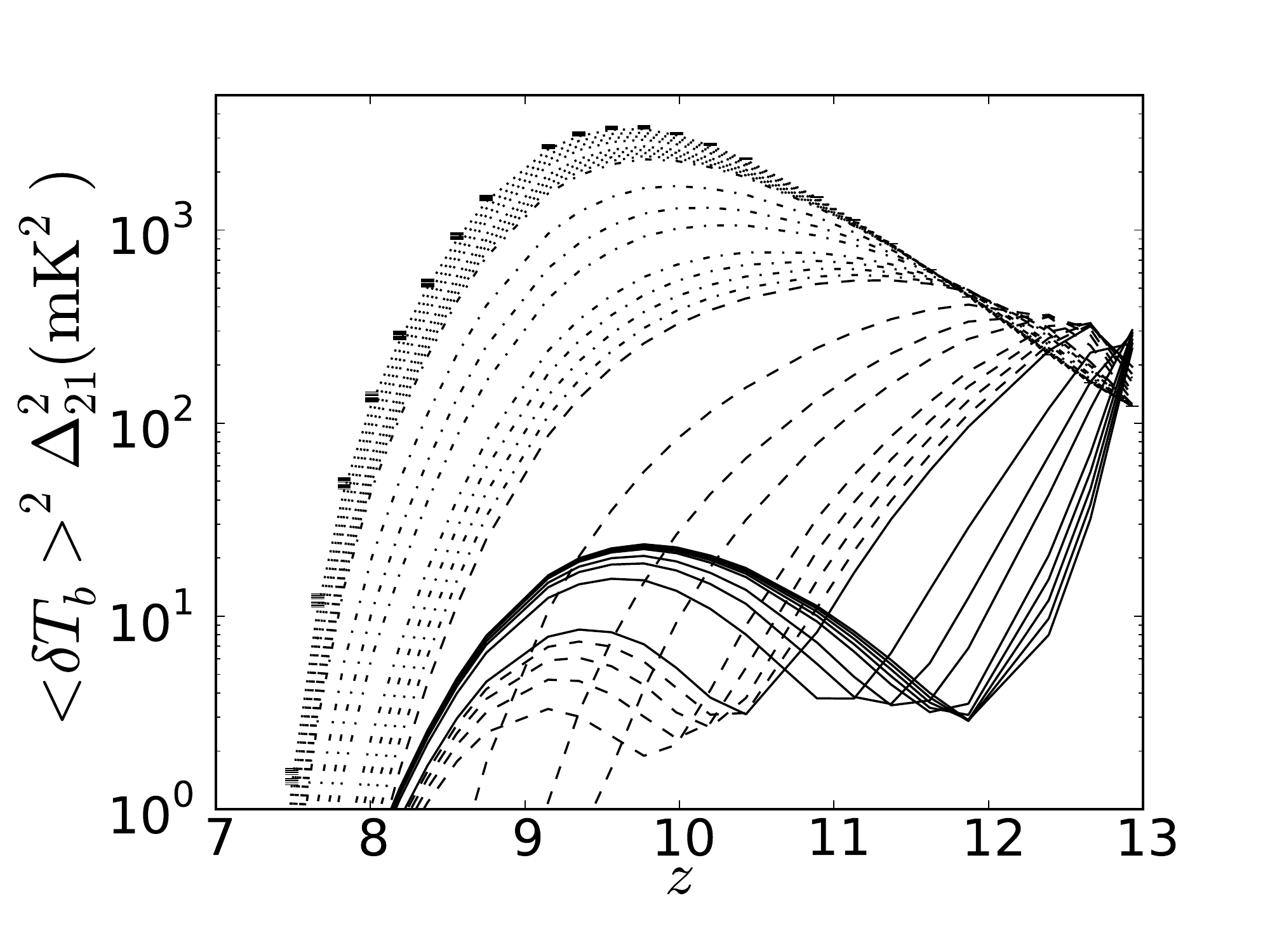}
\caption{\label{monotonicity} Evolution of the 21 cm power spectrum at $k=0.1 \; \rm Mpc^{-1}$ as a function of redshift. Models with  $f_X f_{\star}$ in the ranges $10^{-2}-10^{-1}$, $10^{-3}-10^{-2}$, $10^{-4}-10^{-3}$, and $10^{-5}-10^{-4}$ are plotted in solid lines, dashed lines, dotted-dashed lines, dotted lines, and vertical bars, respectively. If models with $f_X f_{\star} \le 10^{-2}$ are disfavored, the regime of monotonicity, where lowering $f_X f_{\star}$ increases the 21 cm signal, stretches between $z=7$ to $z \sim 11$. However, if we are to include these models and instead disfavor models with extremely low X-ray heating ($f_X f_{\star} \le 10^{-4}$), the regime of monotonicity becomes much narrower, between $z \sim 11$ and $z \sim 12$.}
\end{figure}

In particular, when X-ray heating saturates the 21 cm signal is no longer a monotonic function of $f_X f_{\star}$. Acknowledging this complication, we conducted a study on the monotonicity of the 21 cm signal with respect to X-ray efficiency. Figure \ref{monotonicity} shows the evolution of the 21 cm power spectrum at $k = 0.1 \; \rm Mpc^{-1}$ as a function of redshift. We show models with  $f_X f_{\star}$ in the ranges $10^{-2} - 10^{-1}$, $10^{-3} - 10^{-2}$, $10^{-4}- 10^{-3}$, and $10^{-5} - 10^{-4}$ using solid lines, dashed lines, dotted-dashed lines, dotted lines, and vertical bars, respectively. 

There are three families of curves in Figure \ref{monotonicity}. The largest signals are produced by low X-ray efficiency models ($f_X f_{\star} \le 10^{-3}$), due to the merging of the X-ray and reionization peaks. The second family of curves are those with X-ray efficiencies around $f_X f_{\star} = 10^{-3} - 10^{-2} $, where the signals reflect the shoulder of the X-ray peak. These signals have a large power at $z \sim 11-12$ but quickly decline with decreasing redshift. The third type of curves corresponds to models with $f_X f_{\star} \ge 10^{-2}$. These models descend down their X-ray peaks at $z \ge 12$, and at $z=8-12$ are experiencing another peak due to reionization. The monotonicity problem arises from the fact that at a given redshift, models with higher X-ray heating, due to the reionization peak, can produce signals that is larger than models with lower X-ray heating, which are descending from their X-ray peaks.

If only an upper limit on the 21 cm signal is established, care must be taken when converting this to an upper limit on the X-ray background. One must be sure that the power spectrum lies in the regime of monotonicity. Without any other constraint, there is no regime where the 21 cm signal is completely monotonic. However, if one is able to disfavor X-ray heating models, then one may open up a regime where the 21 cm signal is monotonic. If models with $f_X f_{\star} \le 10^{-2}$ are disfavored, the regime of monotonicity stretches between $z=7-11$. If instead we are to include these models and instead disfavor models with extremely low X-ray heating ( $f_X f_{\star} \le 10^{-4}$), the regime of monotonicity becomes much narrower, occupying the redshifts between $z \sim 11-12$. 

Without disfavoring any models a priori, a robust upper limit on the X-ray background can be obtained by establishing 21 cm signal upper limits at multiple redshifts. With multiple upper limits we can disfavor different $f_X f_{\star}$ models, and find whether the peak at $z=8-13$ is due to reionization, X-ray, or combined X-ray plus reionization. 

Note that this problem is not significant at our redshifts of interest if the upper limit is large (at least in the $\sim 10^2-10^3 \; \rm mK^2$ range). The reionization peak of higher $f_X f_{\star}$ models cannot produce a signal as large as the combined X-ray plus reionization peak exhibited by models of lower $f_X f_{\star}$. In fact, if we only consider upper limits of at least $\sim 10^2 \; \rm mK^2$, the regime of monotonicity stretches between $z=8$ to almost $z=12$.

In addition, note that curves of vastly different X-ray efficiencies in Figure \ref{monotonicity} can intersect. Thereby, unless the detected signal is very large (at least in the $\sim 10^2-10^3 \; \rm mK^2$ range), a single detection is not sufficient to determine the X-ray background. Breaking this degeneracy again requires two or more detections (or at the very least a detection and an upper limit) at different redshifts. With detection at multiple redshifts, one can reconstruct points from Figure \ref{monotonicity} and uniquely infer the X-ray background. 

Our assertion that knowing the redshift of the X-ray peak is sufficient to determine the X-ray background still holds: knowing the position of the peak requires multiple measurements of the power spectrum at and around the peak. Of course, this also implies that knowing the value of the power spectrum at the X-ray peak is sufficient in determining the X-ray background. 

\section{Degeneracies}
Possible sources of degeneracies are the Ly{$\alpha$} and the ionization efficiencies at the signal's redshift. The former is due to the Wouthuysen-Field (WF) mechanism \citep{Wouthuysen} \citep{Field}. The latter stems from the $x_{HI}$ term in equation (1). 

\subsection{Ly$\alpha$-XRB degeneracy}

\begin{figure}
  \centering
   \subfloat[$z=12$]{\label{fa_z12}\includegraphics[scale=0.4]{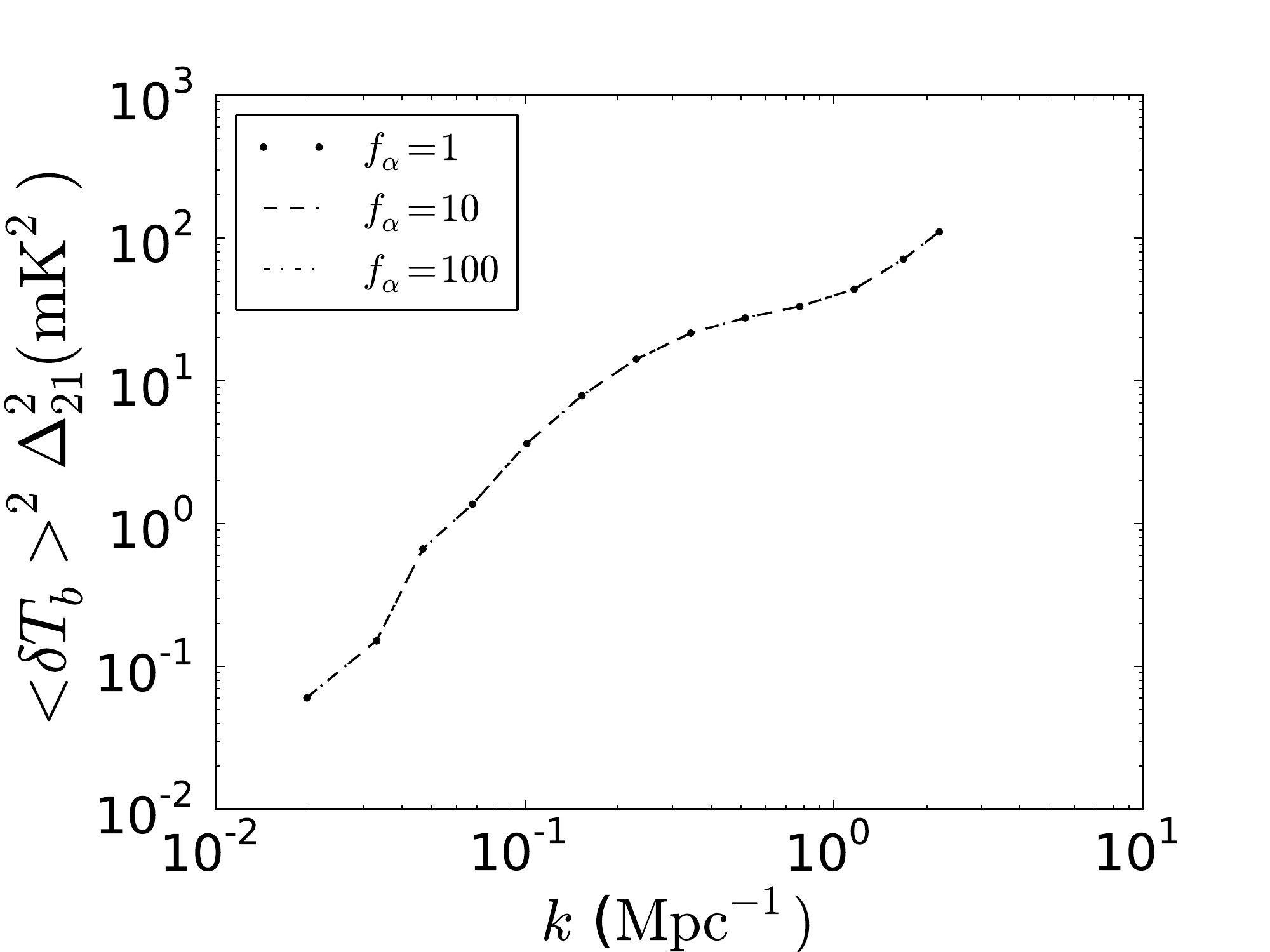}} 
   \subfloat[$z=20$]{\label{fa_z20}\includegraphics[scale=0.4]{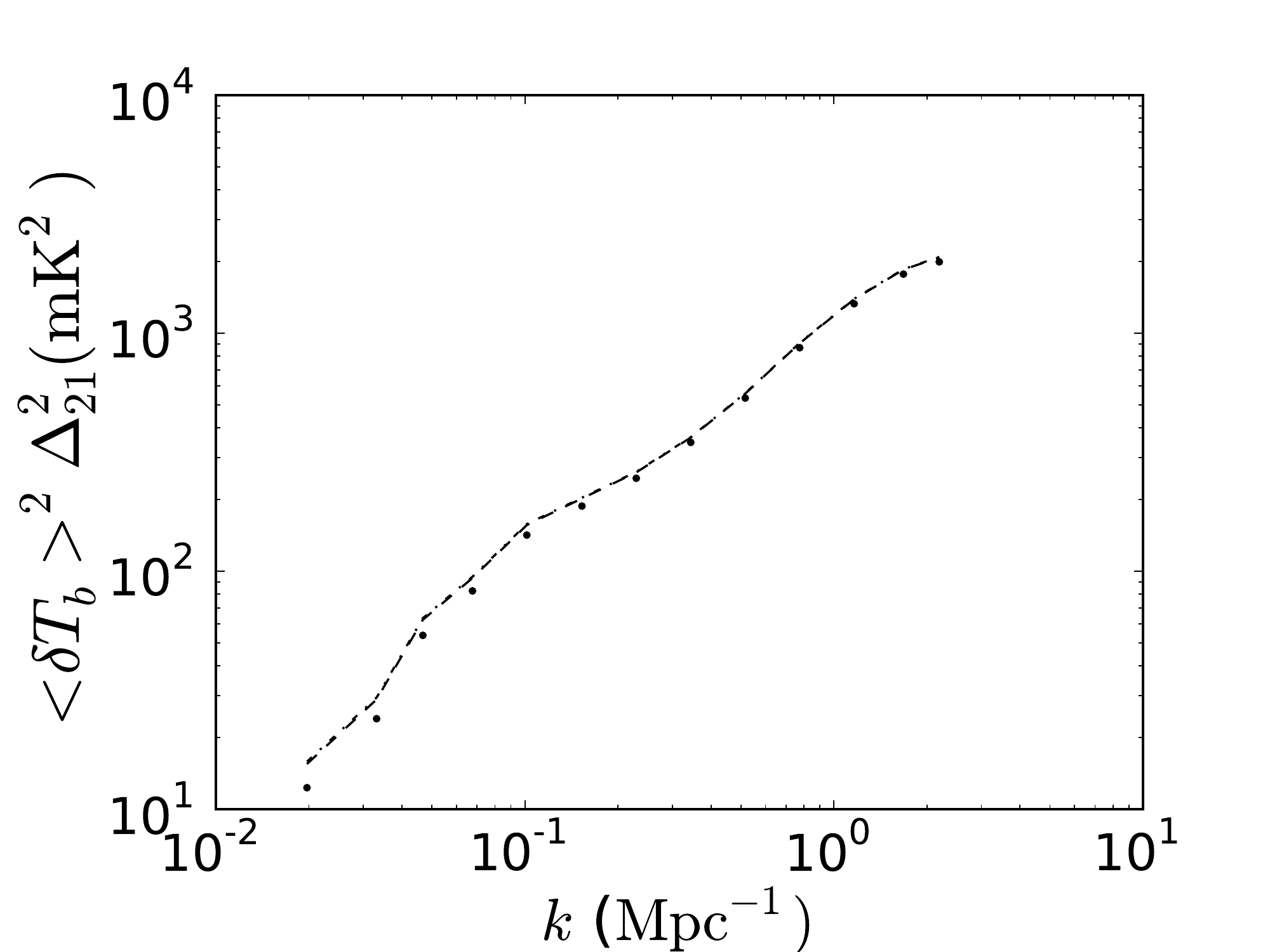}} \\	
   \subfloat[$z=25$]{\label{fa_z25}\includegraphics[scale=0.4]{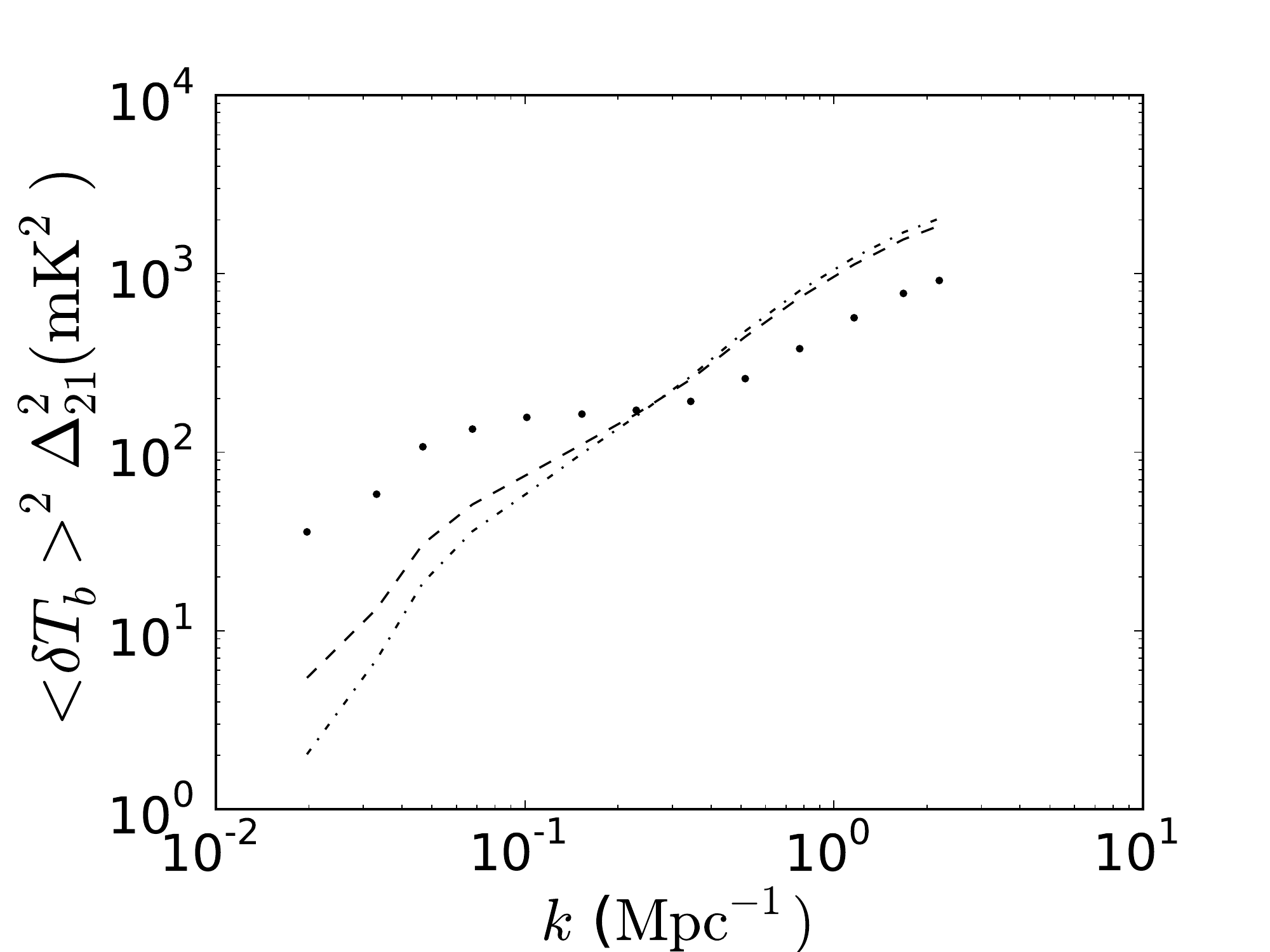}} 	
  \caption{21 cm power spectra at z=10 for various choices of $f_{\alpha}$. A change of $f_{\alpha}$ by two orders of magnitude has a weak influence on the signal at redshifts $z \le 20$. There is considerable degeneracy between $f_X$ and $f_{\alpha}$ at very high redshifts (as seen in the $z=25$ plot) due to the Ly$\alpha$ coupling being unsaturated.}
  \label{falpha}
\end{figure}

The Ly$\alpha$ effect can be parameterized by $f_{\alpha}$, defined as the number of Ly{$\alpha$} photons produced per baryon in stars. In order to explore how the Ly{$\alpha$} efficiency affect on our result, we reran our simulation with different values of $f_{\alpha}$. Our results, plotted in Figure \ref{falpha} show that the 21 cm power spectrum at the redshift of interest ($z \le 15$) is insensitive to changes in the Ly$\alpha$ flux. This insensitivity originates from the fact that the effect of WF coupling on the 21 cm signal is already saturated during the reionization epoch at $z \sim7-15$ \citep{LoebBook}. Thereby, changing the amount of Ly$\alpha$ photons has little effect to the 21 cm intensity. At these redshifts, the degeneracy is lifted due to the 21 cm signal becoming insensitive to variations in the Ly$\alpha$ flux.

\subsection{Ionization efficiency-XRB degeneracy}

\begin{figure}
  \centering
   \subfloat[$z=14$]{\label{fion_z14}\includegraphics[scale=0.4]{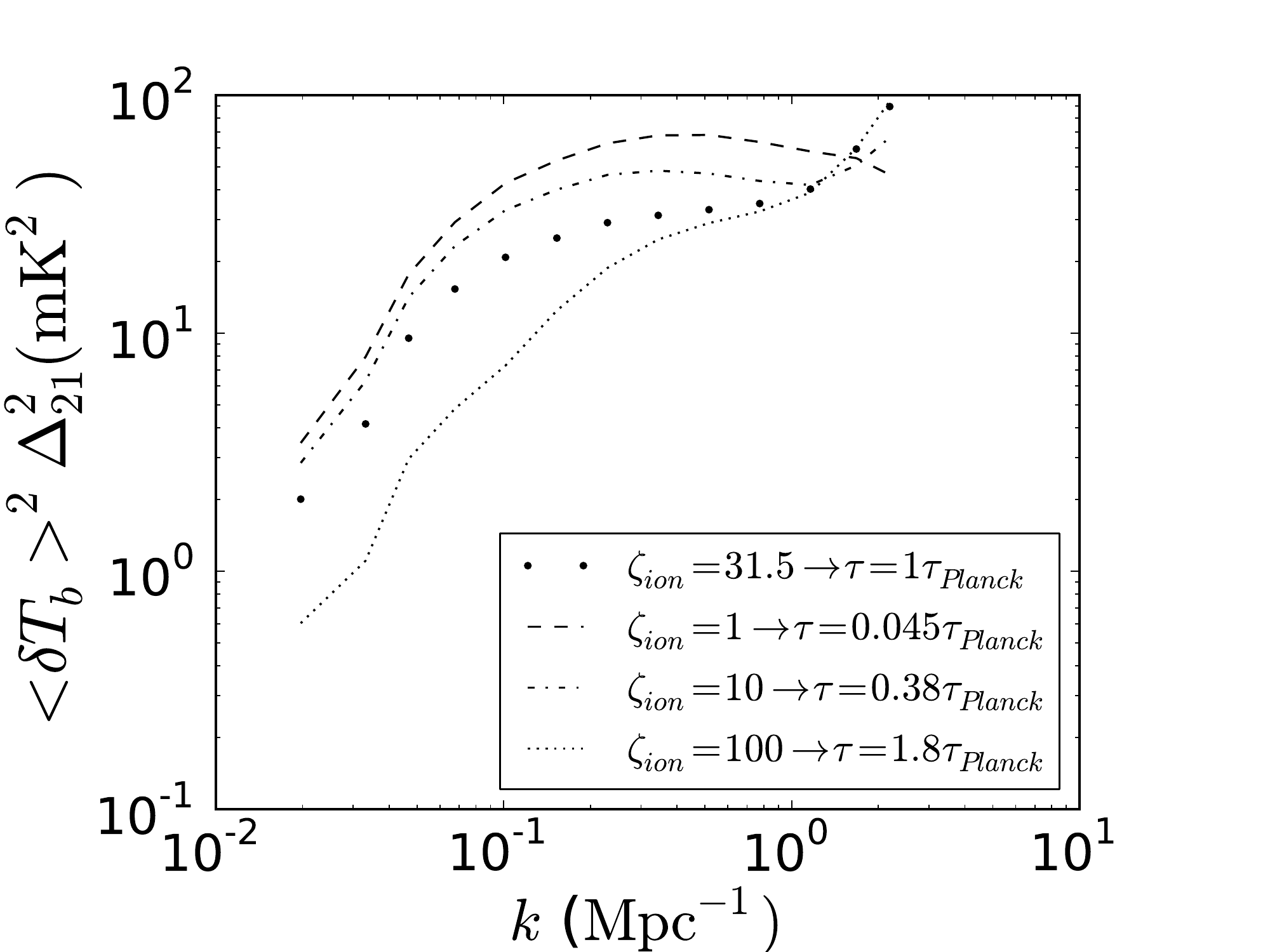}}
   \subfloat[$z=12$]{\label{fion_z12}\includegraphics[scale=0.4]{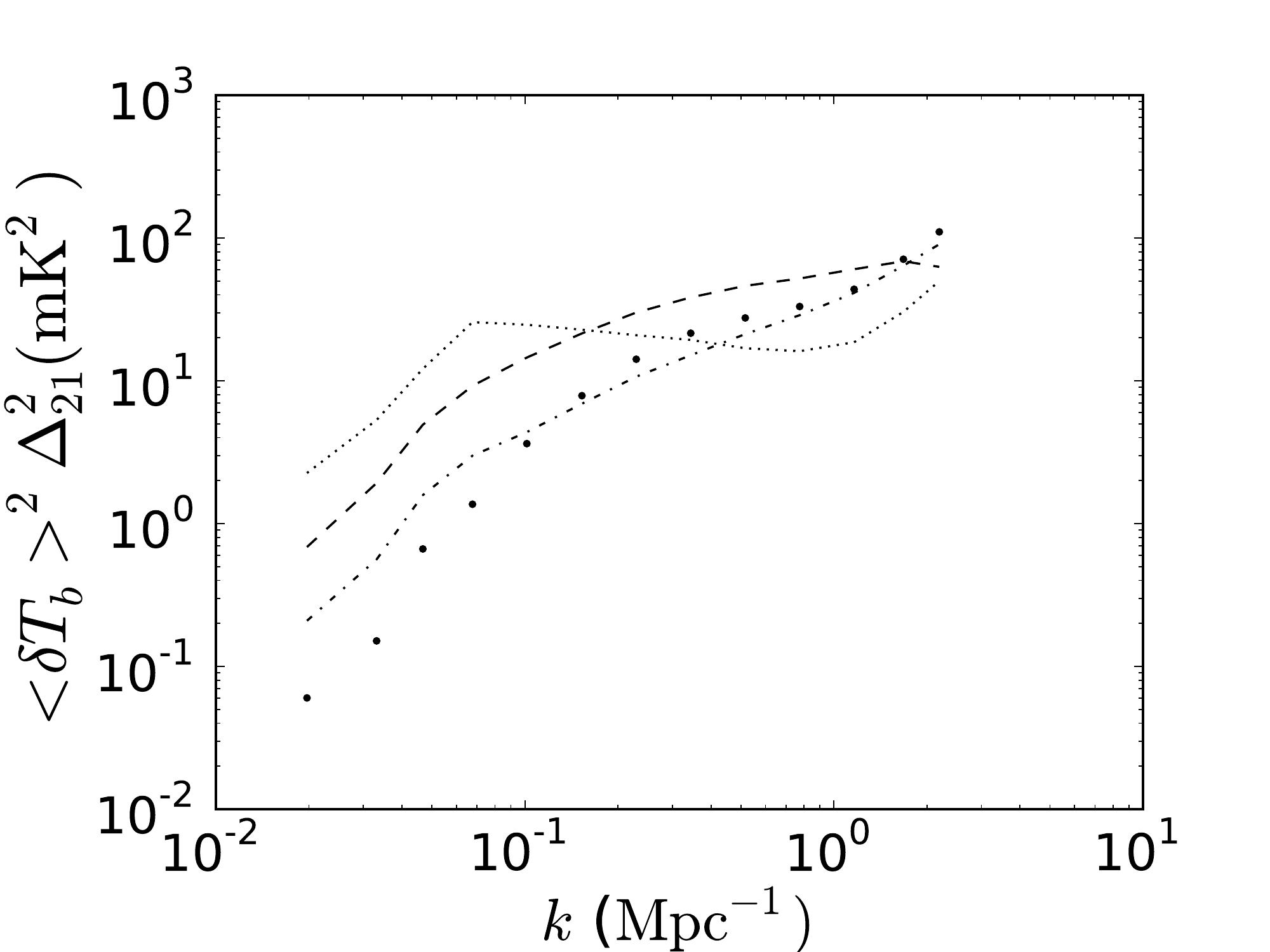}} \\	
   \subfloat[$z=10$]{\label{fion_z10}\includegraphics[scale=0.4]{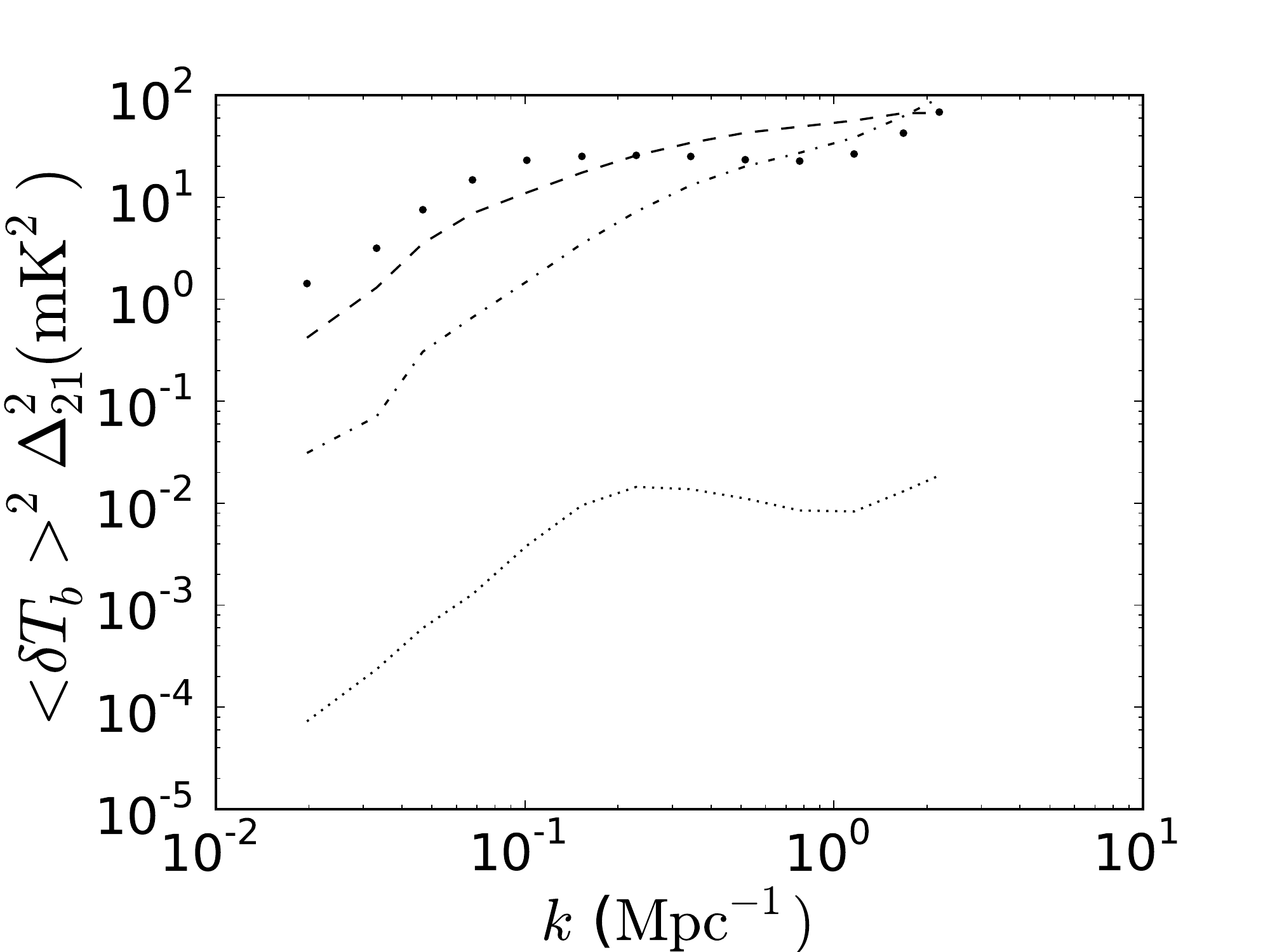}} 	
  \caption{21 cm power spectra at $z=10$ for various choices of $\zeta_{ion}$. The fiducial value of $\zeta_{ion}=31.5$ corresponds to ionization ending at $z\sim7$. The scenarios of $\zeta_{ion}=1$ and  $\zeta_{ion}=10$ failed to fully ionize the universe at $z=6.5$. Another way to showcase this constraint is to calculate the electron scattering optical depth, $\tau$, for each reionization scenario. In the inset of panel (a), the Thomson optical depth $\tau$ of each model is expressed in terms of the value measured by the Planck satellite,
assuming reionization completed by $z=7$.}
  \label{fion}
\end{figure}

The second source of degeneracy involves the ionization efficiency. This effect is implemented in the 21cmFAST code via the parameter $\zeta_{ion}$, where any cell is flagged as being ionized if the collapsed fraction is larger than $\zeta_{ion}^{-1}$ \citep{21cmFAST}. Note that in our prescription, $f_{\star}$ -- the star formation efficiency and $\zeta_{ion}$ --  the ionization efficiency, are independent parameters. While it may appear that changing $f_{\star}$ should change $\zeta_{ion}$, the latter also encodes other unknown parameters (such as the escape fraction of ionizing photons from their host galaxies and the mass function of stars). It is possible therefore for $f_{\star}$ to change while $\zeta_{ion}$ remains constant. The fiducial value of $f_{\star}$ is $31.5$, which is set so that reionization ends at $z \sim 7$. In Figure \ref{fion} we present runs with $\zeta_{ion} = 1, \;10, \; 31.5, \; 100$. 

At first, the results might seem worrying: the changes to the 21 cm signal due to changes in $\zeta_{ion}$ are rather large. However, $\zeta_{ion}$ is also the parameter that determines the neutral fraction of the IGM. Therefore, there is a good constraint on $\zeta_{ion}$ because the end of reionization is constrained by various experiments to be around $z\sim6-7$ \citep{reion}. The end of reionization redshift is very sensitive to changes in $\zeta_{ion}$; our choices of $\zeta_{ion}=1$ and $10$ failed to fully ionize the universe by $z=6.5$ (neutral fraction of $0.94$ and $0.43$ respectively at $z=7$), while $\zeta_{ion}=100$ ended reionization at $z=9.8$. 

We calculate the electron scattering optical depth for each of our models from the equation:
\begin{equation}
\tau(z) = \int_\infty^z (1+z')^3 (1 - X_{HI}) \sigma_{T} \frac {n_{e0} c} {(1+z')H(z)} dz' \; , 
\end{equation}
where $X_{HI}$ is the neutral fraction, $\sigma_T$ is the Thomson cross section, $n_{e0}$ the number density of electrons at $z=0$, and $H(z)$ is given by:
\begin{equation}
H(z) = H_0 (\Omega_M (1+z)^3 + \Omega_\Lambda)^{1/2} \; .
\end{equation}
For each model, we calculate the optical depth at $z=7$ (the redshift when our simulation terminates) and compare it to the Thomson optical depth measured by the Planck mission: $\tau(z=0) = 0.089  \pm 0.032$ \citep{Planck}. In order to calculate the Planck optical depth at $z=7$, we assume a fully ionized universe at $z \le 7$ and subtract the $z=0$ to $z=7$ contribution from the Planck optical depth. For $\zeta_{ion} = 1, 10, 31.5, 100$ we obtain optical depths of $\tau(z=7) = 0.045, \; 0.38, \; 1, \; 1.8 \; \tau_{Planck}(z=7)  $ respectively. 

If we ignore the constraint on $\zeta_{ion}$ from the Planck optical depth, we could still break this degeneracy by noting that modulating $\zeta_{ion}$ affects the power spectrum evolution in a different way than changing $f_X f_{\star}$. Panel (b) of Figure \ref{zevolve} displays the redshift evolution of the 21 cm power spectrum at $k = 0.1 \; \rm Mpc^{-1}$ for different values of $\zeta_{ion}$. Note that changing $\zeta_{ion}$ shifts the location of the reionization peak (the one with the lowest $z$), but keeps the location of the X-ray peak unchanged. As such, if we can locate the position of the X-ray peak, we can use either the peak's redshift or the actual power spectrum at and around the peak to measure the X-ray background free from any degeneracy with $\zeta_{ion}$.

Note that the reionization efficiency uniquely determines the redshift of the reionization peak. One might then naively assume that we could measure the position of the reionization peak and through a similar process uniquely determine $\zeta_{ion}$. However, a low enough X-ray efficiency will merge the X-ray and reionization peaks. Due to the X-ray peak being much larger than the reionization peak, the redshift of this combined peak will correspond more to the X-ray peak's redshift, thereby masking useful information on $\zeta_{ion}$. If one can demonstrate that the peak is solely due to reionization (either because of its low power or by detecting the X-ray peak at a different redshift), one could uniquely determine both $\zeta_{ion}$ and $f_X f_{\star}$, trivially breaking the degeneracy.

Another method to measure the X-ray background free from the $\zeta_{ion}$ degeneracy follows from the result that modulating the reionization efficiency does not change the 21 cm power spectrum beyond the X-ray peak. A robust measurement of the X-ray background can also be conducted at these high redshift ($z \ge 18$ for the fiducial $f_X f_{\star}=0.1$ model), although the foreground would make observations challenging. 

At a sufficiently high redshift, one can also consider the different spatial effects of changing $\zeta_{ion}$ and $f_X f_{\star}$. Due to their short mean free paths, UV photons would only be able to modulate the 21 cm power spectrum at short spatial scales, implying that $\zeta_{ion}$ would have a larger effect on large $k$. 

\section{Effect of changing the X-ray spectrum}

Throughout this paper we have assumed that the X-ray background has a power-law spectrum of the form $L_X \propto (\nu/\nu_0)^{-\alpha}$, where $\nu_0$ is the lowest X-ray frequency present in the IGM and the spectral index $\alpha$ was set to the fiducial value of $1.5$, commonly adopted in the literature \cite{xray_mesinger}.
In this section we investigate the effect of varying the X-ray spectral index, $\alpha$.

\begin{figure}
  \centering
   \subfloat[$f_X f_{\star} = 10^{-1}$]{\label{strongx}\includegraphics[scale=0.4]{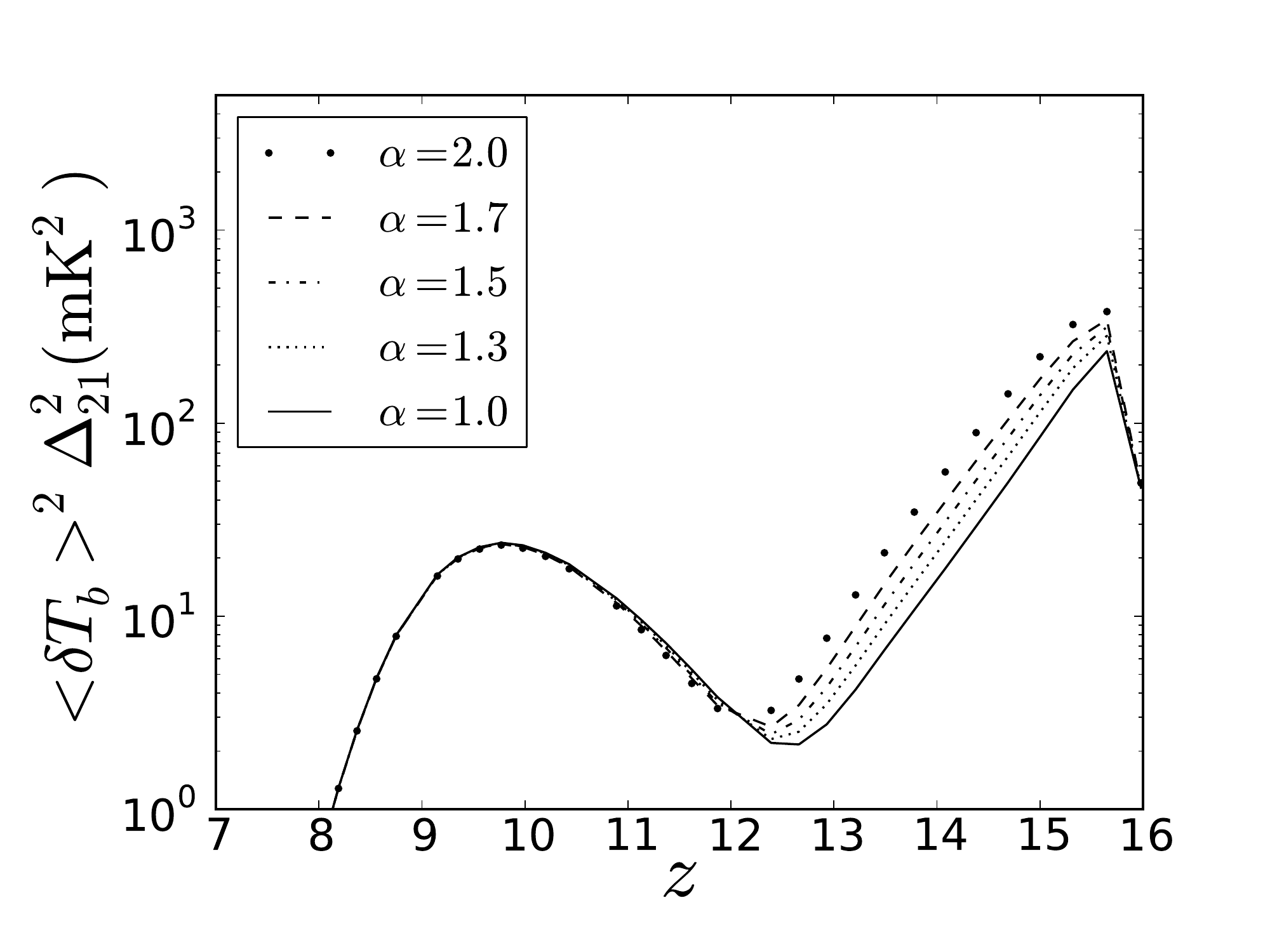}}
  \subfloat[$f_X f_{\star} = 10^{-2}$]{\label{weakx}\includegraphics[scale=0.4]{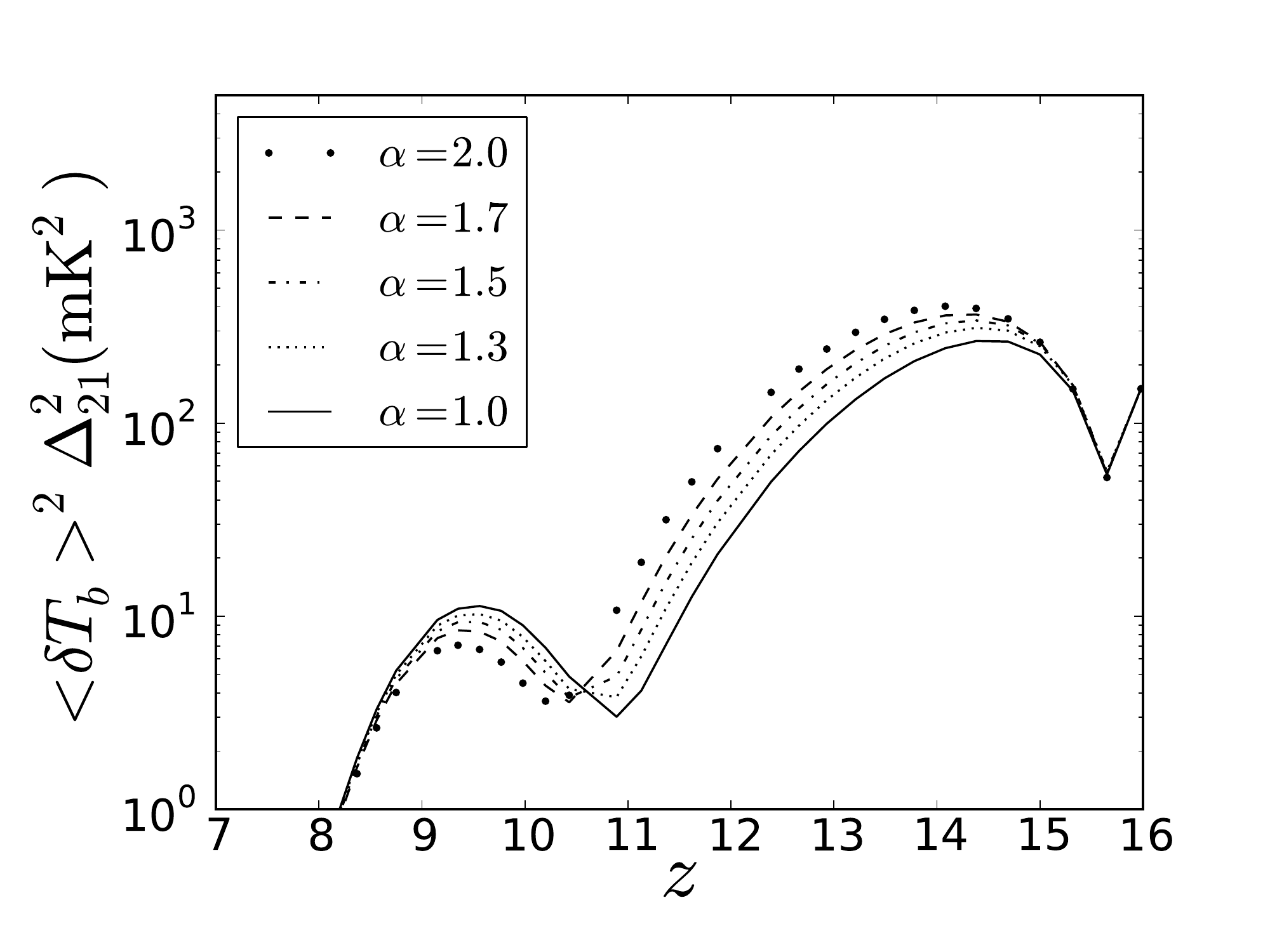}}   
 \caption{The effect of changing the power-law index $\alpha$ of the X-ray spectrum on the $k=0.1 \; \rm Mpc^{-1}$ mode as a function of redshift for models with $f_X f_{\star}=10^{-1}$ and $f_X f_{\star}=10^{-2}$. The effect is not significant at low redshifts where the spin temperature effect saturates, and is modest even at larger redshifts.}  
 \label{spectrum}
\end{figure}

Figure \ref{spectrum} shows the effect of changing $\alpha$ on the $k=0.1 \; \rm Mpc^{-1}$ mode as a function of redshift for models with $f_X f_{\star}=10^{-1}$ and $f_X f_{\star}=10^{-2}$. At redshifts lower than $\sim 12.5$ and $\sim 8.5$ for the $f_X f_{\star}=10^{-1}$ and $f_X f_{\star}=10^{-2}$ models respectively, the effect of spin temperature on the 21 cm signal is saturated, causing the variation in the X-ray spectral index $\alpha$ to have little effect. Between the point of saturation and the X-ray peak the effect is larger, but still rather modest (leading to a difference of at most a factor of $5$ for the spectral indices $\alpha=1$ and $2$). In addition, we have also found that the location of the peaks vary weakly with $\alpha$.

\section{Conclusions}
We have shown that for some values of the X-ray background intensity, first generation 21 cm experiments are capable of detecting the 21 cm signal during the EoR. At $z=10$, an X-ray intensity of $2 \times 10^{-7} \; \rm erg \; \rm sec^{-1} \; \rm cm^{-2} \; \rm steradian^{-1}$ will result in a $10 \sigma$ detection after 1000 hours of MWA observation. 

Figure \ref{zevolve} shows that modulating the X-ray background will shift the position of the X-ray heating peak. Since there is a one-to-one correspondence between the position of this peak and the EoR X-ray background, measuring the signal at multiple redshifts would allow one to measure the X-ray background during the EoR. 

Noting the lack of degeneracies between the Ly$\alpha$ efficiency and $f_X f_{\star}$ at our redshift of interest, as well as the strong constraint on the reionization efficiency (e.g. from measurements of the neutral fraction and the CMB Thomson optical depth) and the fact that modulating $\zeta_{ion}$ and $f_X f_{\star}$ affect the 21 cm power spectrum evolution differently, we have determined that our method is robust.

If the reionization peak is located, one can use its redshift to measure the ionization efficiency (or equivalently the Thomson optical depth) of the IGM. This will help in inferring the reionization history of the universe independent from other constraints (e.g. quasar spectra or CMB electron scattering optical depth). 

\bigskip
\paragraph*{Acknowledgments.}

We thank Andrei Mesinger for his creation and maintenance of the 21cmFAST code. We would also like to thank Judd Bowman, Adam Beardsley, and Aaron Ewall-Wice for helpful discussions. This work was supported in part by NSF grant AST-0907890 and NASA grants NNX08ALY36 and NNA09DB30A.

\bibliography{bibliography}

\end{document}